\documentclass[10pt,twocolumn]{article}
\usepackage[top=1.8cm, bottom=1.8cm, left=1.8cm, right=1.8cm]{geometry}
\usepackage{helvet}  %
\usepackage{courier}  %
\usepackage[hyphens]{url}  %
\usepackage{graphicx}  %
\usepackage{pifont}
\usepackage{stackengine}
\usepackage{amsmath}
\usepackage{amssymb} %

\usepackage{amsmath}
\usepackage[hang,flushmargin,stable]{footmisc}
\usepackage{graphicx}
\graphicspath{ {figures/} }

\usepackage{color}
\usepackage{afterpage}

\usepackage{xcolor}
\definecolor{light-gray}{gray}{0.95}

\usepackage{listings}
\usepackage[small,bf]{caption}
\DeclareCaptionFont{white}{\color{white}}
\DeclareCaptionFormat{listing}{\colorbox[cmyk]{0.43, 0.35, 0.35,0.01}{\parbox{\textwidth}{\hspace{15pt}#1#2#3}}}
\usepackage{etoolbox}

\makeatletter
\patchcmd{\ttlh@hang}{\parindent\z@}{\parindent\z@\leavevmode}{}{}
\patchcmd{\ttlh@hang}{\noindent}{}{}{}
\makeatother

\usepackage{indentfirst}
\usepackage{url}
\usepackage{subcaption}
\usepackage{xspace}
\usepackage{multirow}
\definecolor{darkred}{RGB}{120,0,30}
\newcommand{\edit}[1]{#1}

\usepackage{threeparttable}
\usepackage{tikz}
\usepackage{bbding}
\usepackage[all]{xy}
\usepackage{algorithm}
\usepackage[noend]{algorithmic}

\usepackage{array}
\usepackage{multirow}
\usepackage{paralist}
\usepackage{url}
\usepackage{graphicx}
\usepackage{latexsym}
\usepackage{amssymb}
\usepackage{amsfonts}
\usepackage{psfrag}
\usepackage{wrapfig}
\usepackage{comment}

\DeclareRobustCommand{\ttfamily}{\fontfamily{lmtt}\selectfont}

\usepackage{xcolor}
\usepackage{booktabs}
\usepackage{graphicx}
\usepackage{paralist}
\usepackage{times}
\renewcommand{\footnotesize}{\fontsize{8}{9}\selectfont}
\usepackage[compact]{titlesec}
\titlespacing*{\section}{0pt}{*4}{4pt}
\titlespacing{\subsection}{0pt}{*2}{2pt}
\titlespacing{\subsubsection}{0pt}{*2}{2pt}

\definecolor{darkgreen}{RGB}{130,0,0}
\definecolor{darkblue}{RGB}{57,79,99}
\usepackage[bookmarks=false, colorlinks=true, plainpages = false,  linkcolor = darkgreen,   citecolor = darkgreen, urlcolor = darkblue, filecolor = blue]{hyperref}
\usepackage[hyphenbreaks]{breakurl}
\PassOptionsToPackage{hyphens}{url}\usepackage{hyperref}

\let\oldbibliography\thebibliography
\renewcommand{\thebibliography}[1]{%
  \oldbibliography{#1}%
  \setlength{\itemsep}{0pt}%
}

\renewcommand{\scriptsize}{\fontsize{7.5}{9.5}\selectfont}

\makeatletter
\def\url@leostyle{%
  \@ifundefined{selectfont}{\def\UrlFont{\small\ttfamily}}%
  {\def\UrlFont{\scriptsize\ttfamily}}%
}
\makeatother	
\newcommand{\descr}[1]{\vspace{0.075cm} \noindent \textbf{#1}}
\newcommand{\descrit}[1]{\vspace{0.05cm} \noindent \emph{#1}}

\newif\ifcomment
\commenttrue 
\ifcomment
	\newcommand{\edc}[1]{\textbf{\em\color{red}#1}}
    \newcommand{\luca}[1]{\textbf{\em\color{blue}#1}}    
\else  
    \newcommand\edc[1]{}
    \newcommand\luca[1]{}
\fi

\renewcommand{\footnoterule}{%
  \kern -3pt
  \hrule width 1in
  \kern 2pt
}

\makeatletter
\def\url@leostyle{%
  \@ifundefined{selectfont}{\def\UrlFont{}}%
  {\def\UrlFont{}}%
}
\makeatother
\usepackage[hyphenbreaks]{breakurl}

\usepackage{hhline}

\newcommand{\oto}{{O2O}\xspace}
\newcommand{\iptip}{{IP2IP}\xspace}
\newcommand{\iptiplus}{{IP2IP+intersection}\xspace}
\newcommand{\STA}{{STA}\xspace}

\begin{document}

\sloppy 

\title{\bf On Collaborative Predictive Blacklisting\thanks{A preliminary version of this paper appears in ACM SIGCOMM's Computer Communication Review. This is the full version.}}

\author{Luca Melis, Apostolos Pyrgelis, Emiliano De Cristofaro\\[0.5ex]
{\normalsize University College London}}

\date{}

\maketitle

\begin{abstract}
Collaborative predictive blacklisting (CPB) allows to forecast future attack sources based on logs and alerts contributed by multiple organizations. Unfortunately, however, research on CPB has only focused on increasing the number of predicted attacks but has not considered the impact on false positives and false negatives. Moreover, sharing alerts is often hindered by confidentiality, trust, and liability issues, which motivates the need for privacy-preserving approaches to the problem.
In this paper, we present a measurement study of state-of-the-art CPB techniques, aiming to shed light on the actual impact of collaboration. To this end, we reproduce and measure two systems: a non privacy-friendly one that uses a trusted coordinating party with access to all alerts~\cite{soldo2010predictive} and a peer-to-peer one using privacy-preserving data sharing~\cite{freudiger2015controlled}. We show that, while collaboration boosts the number of predicted attacks, it also yields high false positives, ultimately leading to poor accuracy.
This motivates us to present a hybrid approach, using a semi-trusted central entity, aiming to increase utility from collaboration while, at the same time, limiting information disclosure and false positives. This leads to a better trade-off of true and false positive rates, while at the same time addressing privacy concerns.
\end{abstract}

\section{Introduction}
\label{section:introduction}
Filtering connections from/to malicious hosts is often used to reduce network attacks and their impact. Due to the impossibility of performing expensive computations in real-time on each connection, filtering is usually done via simple look-ups, using periodically updated lists of suspicious hosts, i.e., {\em blacklists}. 
These can be created locally and/or by obtaining the most prolific attack sources from alert repositories such as DShield.org or DeepSight~\cite{sym06}.

In~\cite{katti2005collaborating}, Katti et al.~study the prevalence of ``correlated'' attacks, i.e., mounted by the same sources against different networks. They find them to be very common, and highly targeted, suggesting that real-time collaboration between victims could improve malicious IP detection time. 
Zhang et al.~\cite{zhang2008highly} are the first to introduce the concept of {\em collaborative predictive blacklisting} (CPB): different organizations send their logs to a central authority that, in turn, provides them with customized blacklists based on relevance ranking. In follow-up work, Soldo et al.~\cite{soldo2010predictive} improve on~\cite{zhang2008highly} by replacing ranking with an implicit recommender system.
Overall, collaborative approaches to threat mitigation are increasingly advocated, with more and more efforts to promote information sharing, including those proposed by CERT~\cite{cert}, RedSky Alliance~\cite{redsky}, Facebook's ThreatExchange~\cite{fb}, or the White House~\cite{whitehouse_last}.

In this work, we focus on two open problems that remain largely unaddressed w.r.t.~the impact of collaboration on (1) false positives/negatives, and (2) privacy.
Prior work on CPB~\cite{soldo2010predictive,zhang2008highly} only focuses on measuring ``hit counts'', i.e., the number of true positives, but fails to account for incorrect predictions---i.e., false positive/negatives.
Moreover, real-world deployment of collaborative blacklisting is hindered by confidentiality issues, as well as trust, liability, and competitiveness concerns as sharing alerts could harm an organization's reputation or disclose sensitive information about customers and business practices~\cite{CSRIC}. To the best of our knowledge, the peer-to-peer model proposed by Freudiger et al.~\cite{freudiger2015controlled} is the only privacy-friendly approach to the problem:
organizations interact in a pairwise manner, aiming to privately estimate the benefits of collaboration, and then share data with ``good'' partners. %
However, as discussed later in this paper, it is not clear how to deploy their decentralized techniques in practice.

First, we reproduce, measure, and compare the centralized (non-private) system by Soldo et al.~\cite{soldo2010predictive} vs the peer-to-peer privacy-friendly one by Freudiger et al.~\cite{freudiger2015controlled}, using alerts obtained from DShield.org, involving 70 organizations which report an average of 4,000 daily events over a 15-day time window. 
We finding that the former~\cite{soldo2010predictive} achieves high hit counts (almost doubling correct predictions compared to no collaboration), but its F1 accuracy is ultimately poor ($14\%$) due to high false positives. Whereas, the latter~\cite{freudiger2015controlled} allows for better control over incorrect predictions, thus resulting in a better F1 score overall ($29\%$), but actually only slightly improves the hit counts over no collaboration since its peer-to-peer approach limits the amount of data that gets shared.

\edit{Our measurements lead to the intuition that, if one needs to control false positives, a controlled data sharing approach might kill two birds with one stone: (1) help organizations find a better trade-off between prediction improvement and increase in false positives, and (2) do so while actually minimizing exposure of possibly confidential data.}
Therefore, we introduce and analyze a novel hybrid model, relying on a semi-trusted authority, or \STA, which acts as a coordinating entity to facilitate clustering without having access to the raw data. The \STA clusters contributors based on the similarity of their logs (without seeing these logs), and helps organizations in the same cluster to share relevant data. Toward this goal, we perform a set of measurements to shed light on (i) how to cluster organizations, (ii) what should be shared among them, and (iii) how to measure the effect of collaboration on accuracy. 

We experiment with a few clustering algorithms using the number of common attacks as a measure of similarity, which can be computed in a privacy-preserving way, and experiment with privacy-friendly within-clusters sharing strategies, namely, only disclosing the details of common/correlated attacks.
Overall, we show that our new hybrid model outperforms~\cite{freudiger2015controlled} in terms of hit counts (4x), while achieving better accuracy than~\cite{soldo2010predictive} (2x).

\section{Preliminaries}
\label{sec:preliminaries}

\subsection{Datasets}\label{subsec:dataset}
We gather a dataset of blacklisted IP addresses from DShield.org, a collaborative firewall log correlation system to which various organizations volunteer daily alerts. Each entry in the logs includes a pseudonymized {\em Contributor ID} (the target), {\em source IP} address (the attacker), {\em source} and {\em target port} number, and a {\em timestamp}. An example of an entry log is illustrated in Table~\ref{table:DShield data}. 

\begin{table}[t]
\centering
\resizebox{0.99\columnwidth}{!}{
\begin{tabular}{| c | c | c | c | c | }
\hline
Contributor ID & Source ID & Source Port & Target Port & Timestamp \\ 
\hline
...D982918 & 104.217.230.059 & 6000 & 1433 & 2015-06-06 11:49:32 \\ 
\hline
\end{tabular}}
\vspace{-0.2cm}
\caption{Example of an entry in the DShield logs.}
\label{table:DShield data}
\vspace{-0.2cm}
\end{table}

With DShield's permission, we collect logs using a web crawler, from February to September 2015, 
gathering, on average, 10 million logs from 120,000 organizations every day.
We exclude entries for invalid or non-routable IP addresses, and discard port numbers, then, for each IP address, we extract its /24 subnet and use /24 addresses for all experiments, following experimental choices made in prior work~\cite{freudiger2015controlled,soldo2010predictive,zhang2008highly}. 
This does not necessarily mean that predictive blacklisting algorithms will blacklist entire /24 subnets, since blacklisting an address does not imply blocking all its traffic, but rather subject it to further scrutiny, e.g., enforcing rate limiting or only allowing outgoing packets. Nonetheless, recall that our main goal here is to compare the impact of different collaboration approaches on prediction.

We select a 15-day period, May 17--31, 2015 and restrict our evaluations to a reasonably-sized sample of regularly contributing organizations. We select the top-100 contributors, based on the number of unique IPs reported, that also report logs every day during the 15 days and notice that most contributors (around 60) submit less than 100K logs, while fewer (around 20) submit between 
100K and 500K, and only a few organizations contribute large amounts of logs (above 1M). Then, we pick 70 organizations, for each time window, leaving out the top-10 and the bottom-20 contributors. %
\edit{We do so, like in previous work~\cite{freudiger2015controlled,soldo2010predictive}, to minimize bias.
More specifically, the top contributors contribute a huge number of IPs (order of magnitudes more than other contributors) which might be irrelevant to most organizations, whereas, the bottom ones only report very few logs, thus adding little or nothing to the collaboration.}
Our final sample dataset includes 30 million attacks, contributed by 118 different organizations over 15 days, each reporting a daily average of 600 suspicious (unique) IPs and 4,000 attack events.
\edit{This constitutes our ``ground truth'': if an IP appears in the blacklist for an organization, it is considered to be malicious for that organization.}

We use this dataset both as {\em training} and {\em testing} sets -- more precisely, we consider a sliding window of 5 days for training and 1 day for testing, as done in previous work~\cite{freudiger2015controlled,soldo2010predictive}. 

Note that we have also repeated our experiments %
on two more sets of DShield logs, using another 15-day periods (over Feb-Dec 2015), but have not found any significant difference in the results. %

\descr{Notation.} %
We use notation $\mathcal{O} = \{O_{i}\}_{i=1}^{n}$ to denote a group of $n$ organizations, where each $O_i$ holds a dataset $D_i$ of alerts, i.e., suspicious IP addresses along with the related timestamp.
We aim to predict IP addresses generating attacks to each $O_i$ in the next day, using, as the training set, both its local dataset $D_i$, as well the set $D'_{i}$, with suspicious IP addresses obtained by collaborating with other organizations. As discussed above, %
we consider $n=70$ organizations using alerts collected from DShield.

\subsection{EWMA Time Series Prediction}\label{subsec:ewma}
We use Exponentially Weighted Moving Average (EWMA)  to perform prediction. 
Given a signal over time $r(t)$, we indicate with $\tilde{r}(t + 1)$ the predicted value of $ r (t + 1) $, given past observations $r(t')$ at time $ t' \leq t$. The predicted signal is computed as:\vspace{-0.35cm}
\begin{equation}
\label{eq:ewma}
\tilde{r} (t + 1) = \sum_{t' = 1}^{t} \alpha \cdot (1 - \alpha)^{t - t'} \cdot r(t')  \vspace{-0.25cm}
\end{equation}
where $ \alpha \in (0, 1) $ is a smoothing coefficient, $ t' = 1, \ldots, t $ denotes the training window, and $t+1$ is the time slot to be predicted. For small values of $ \alpha $, EWMA aggregates past information uniformly across the training window, while, with a large $ \alpha $, the prediction algorithm focuses more on events taking place in the recent past.

\subsection{Metrics}\label{subsec:metrics}
Throughout our evaluations, we use the following metrics to evaluate the 
performance of the predictions.

\descr{True and False Positives.} For each time window and for each organization, we count \textit{True Positives} (TP) as well as \textit{False Positives} (FP). A TP occurs when the prediction algorithm includes an IP address in an organization's predictive blacklist that does appear in its testing set, and a FP -- when it does not.

\descr{False Negatives.} %
For each time window/organization, we generate predictive \textit{whitelists}, i.e., sets of IPs that are not likely to attack an organization the next day, and %
count a \textit{False Negative} (FN) when a whitelisted IP address instead appears in the testing set.

\descr{TP Improvement and FP/FN Increase.} We also measure the average improvement/increase in TP, FP, and FN when compared to a baseline local approach, i.e., when no collaboration occurs between organizations and each of them makes its predictions based only on its local dataset. The improvement in TP is calculated as: 
$TP_{impr}=(TP_{c} - TP)/TP$
where $ TP_c$ is the number of true positives after collaboration and TP without. 
Similarly, the increase in FP and FN is denoted, resp., as
$FP_{incr} = (FP_{c} - FP)/FP$ and $FN_{incr} = (FN_{c} - FN)/FN$.

\descr{Precision, Recall and F1-Score.} We calculate the True Positive Rate (TPR), aka 
{\em recall}, False Positive Rate (FPR), as well as Positive Predictive Value (PPV), aka 
{\em precision}, defined as: TPR=TP/(TP+FN), FPR=FP/(FP + TN), PPV=TP/(TP + FP),
and derive the F1 measure, i.e., \vspace{-0.15cm}
{\small $$ \text{F1}=2\cdot\frac{\text{PPV} \cdot \text{TPR}}{\text{PPV}+\text{TPR}}\vspace{-0.15cm}$$} %

\descr{\bf\em Remarks on FP:} The absence of an IP from our testing set can occur {\em either} when the IP is not considered suspicious {\em or} if it does not generate requests.
While we cannot actually distinguish between the two cases, in the latter a FP is actually less ``severe'' than in the former, thus our FP count may be a bit more conservative.
However, our main goal is really to measure and compare with each other the impact of {\em different} collaboration strategies on predictions so we use this method without loss of generality.

\begin{figure*}[t]
\centering
\begin{subfigure}[t]{0.325\textwidth}
\centering
\includegraphics[width=1\textwidth,height=0.18\textheight]{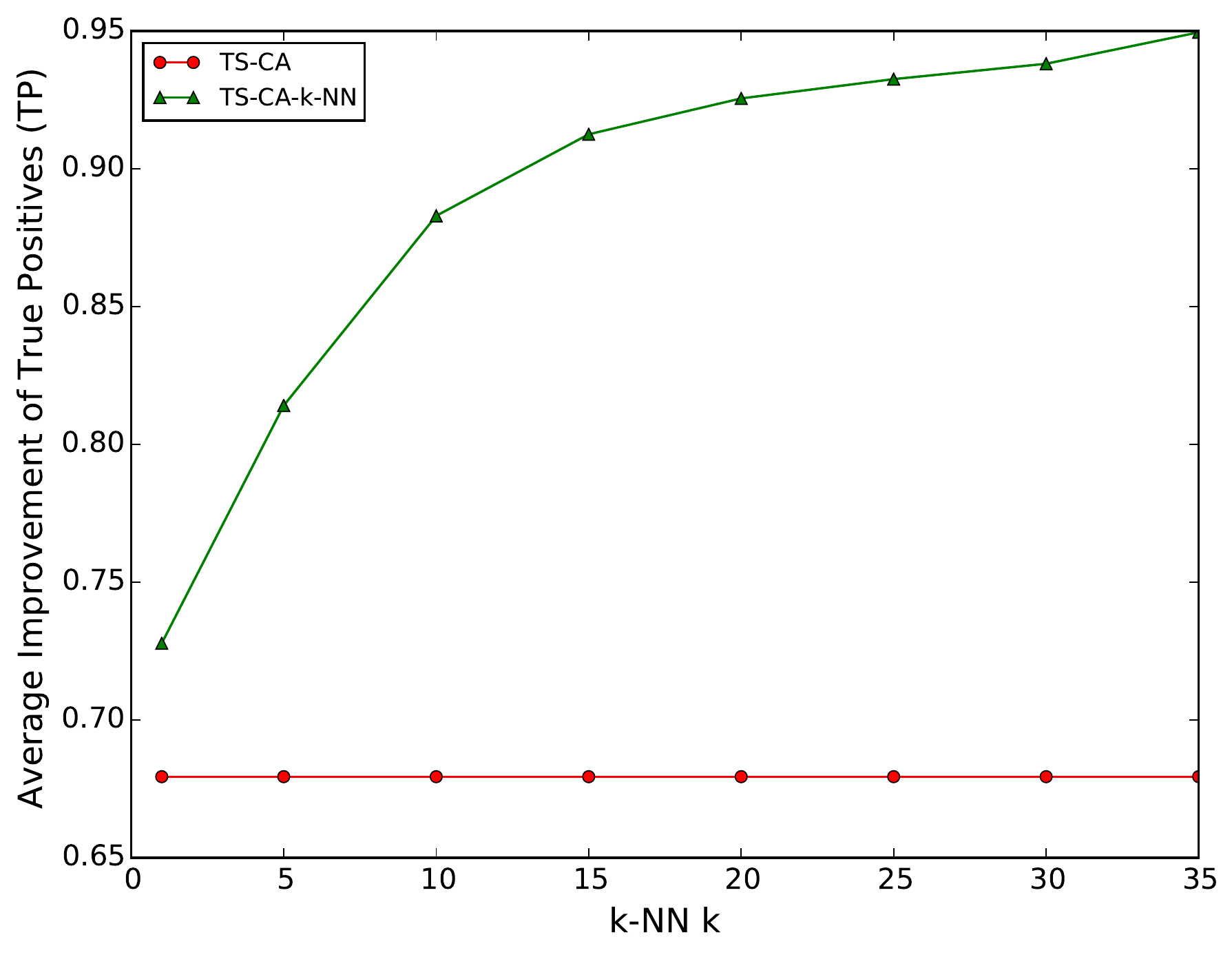}
\caption{ %
\label{fig:soldo-tp-incr}}
\end{subfigure}
\begin{subfigure}[t]{0.325\textwidth}
\centering
\includegraphics[width=1\textwidth,height=0.18\textheight]{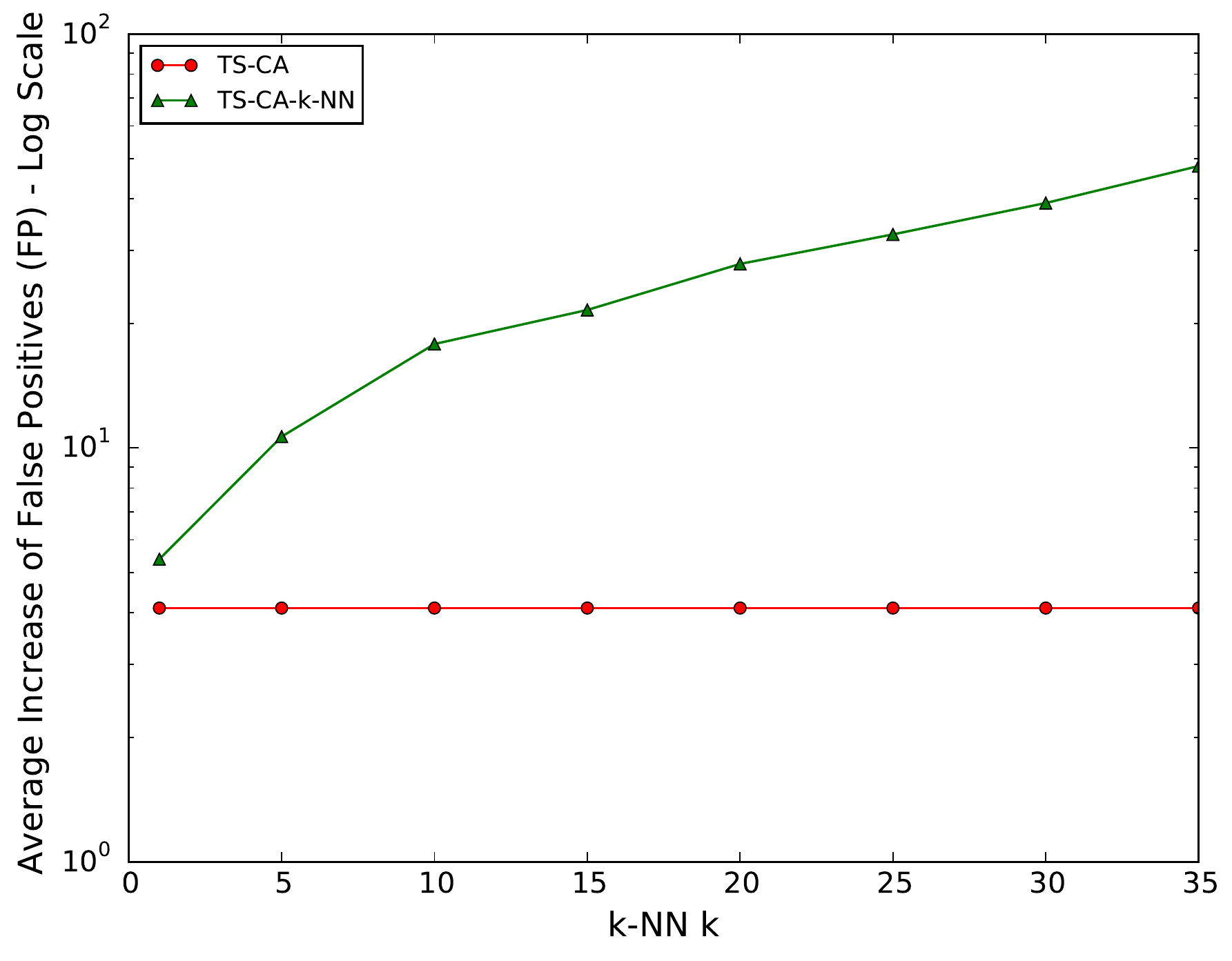}
\caption{ %
\label{fig:soldo-fp-incr}}
\end{subfigure}
\begin{subfigure}[t]{0.325\textwidth}
\centering
\includegraphics[width=1\textwidth,height=0.18\textheight]{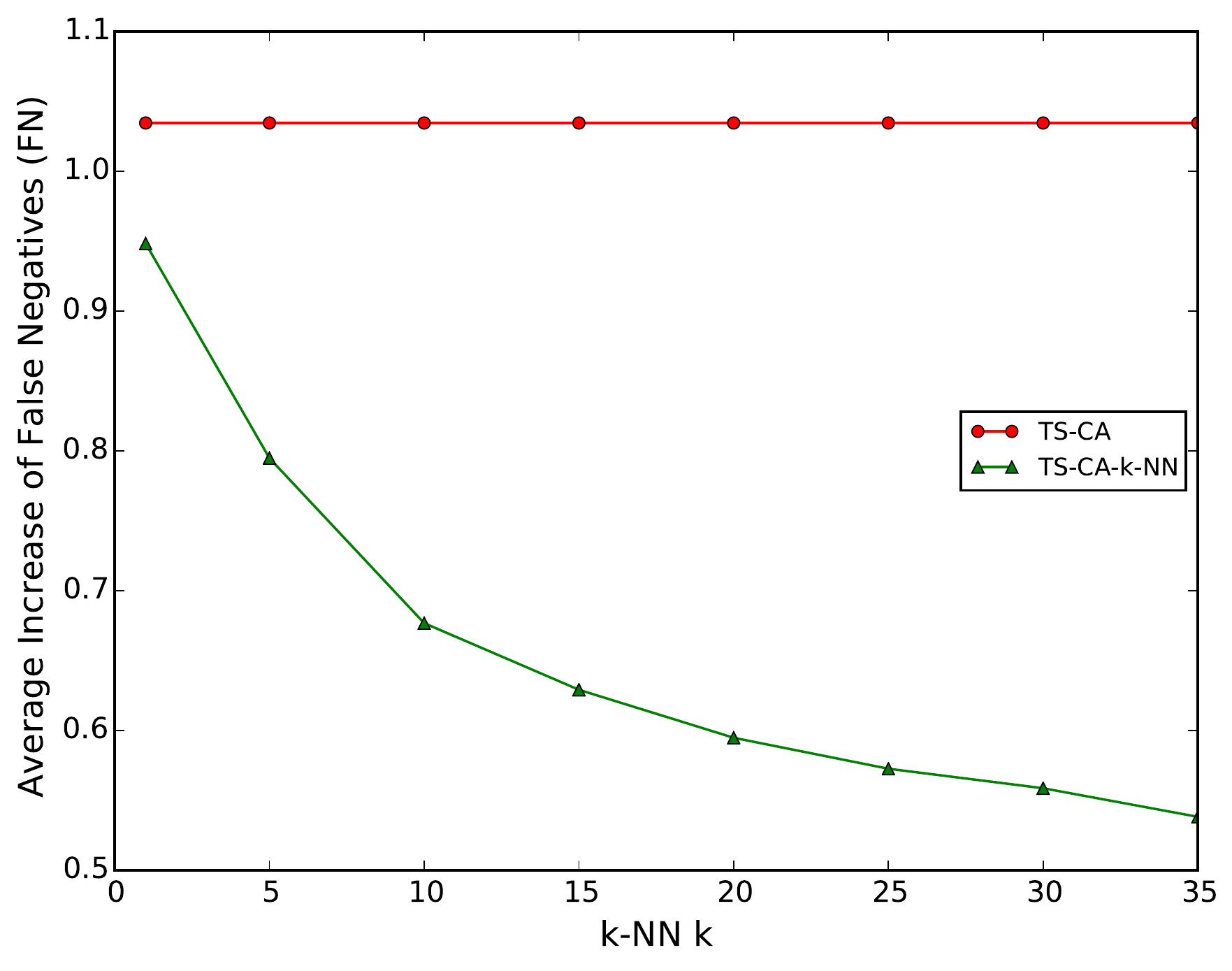}
\caption{ %
\label{fig:soldo-fn-incr}}
\end{subfigure}
\begin{subfigure}[t]{0.325\textwidth}
\centering
\includegraphics[width=1\textwidth,height=0.18\textheight]{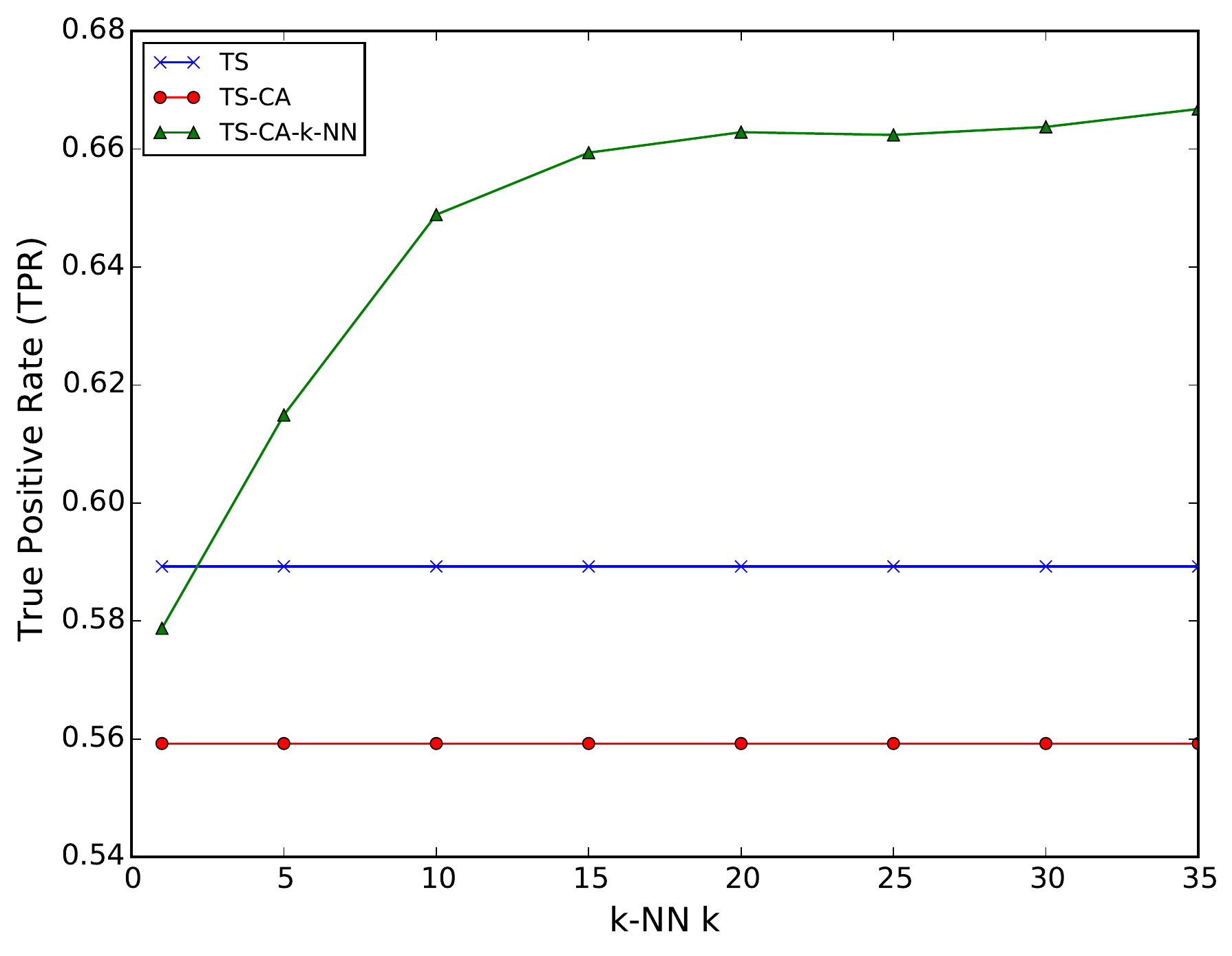}
\caption{ %
\label{fig:soldo-tpr}}
\end{subfigure} 
\begin{subfigure}[t]{0.325\textwidth}
\centering
\includegraphics[width=1\textwidth,height=0.18\textheight]{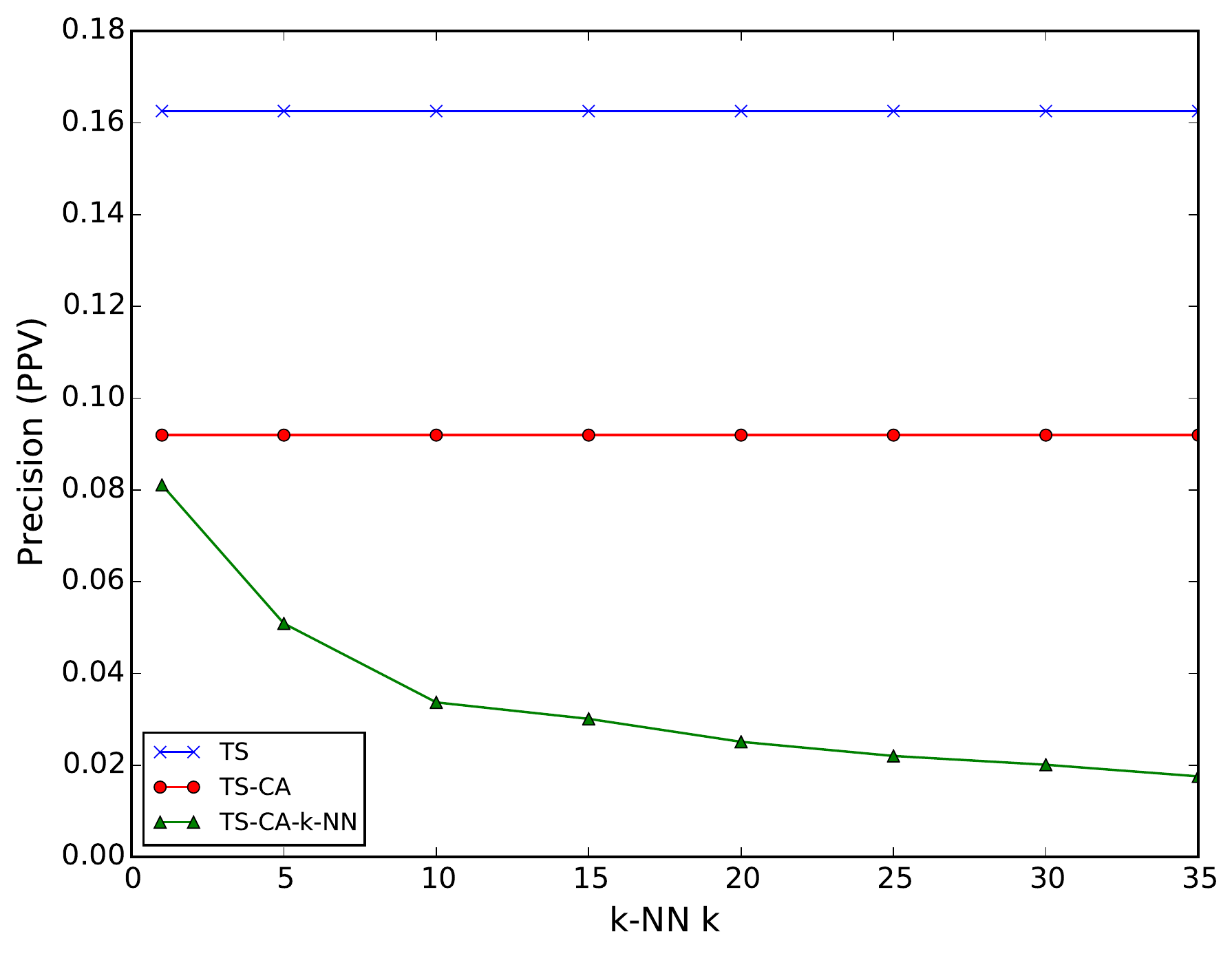}
\caption{ %
\label{fig:soldo-ppv}}
\end{subfigure} 
\begin{subfigure}[t]{0.325\textwidth}
\centering
\includegraphics[width=1\textwidth,height=0.18\textheight]{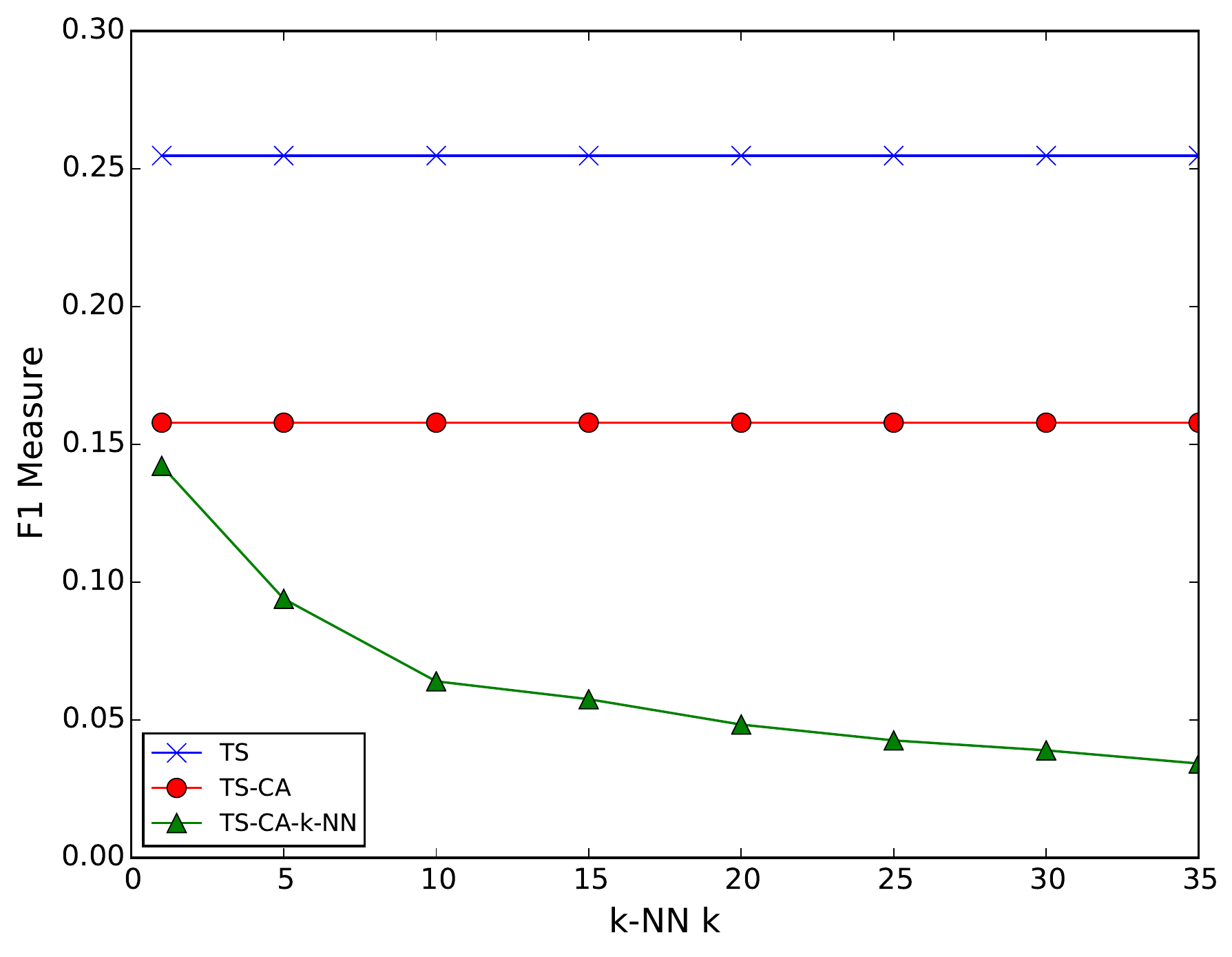}
\caption{ %
\label{fig:soldo-f1}}
\end{subfigure}
\vspace{-0.15cm}
\caption{Soldo et al.~\cite{soldo2010predictive}: (a) TP improvement, (b) FP increase (y-axis in log scale), (c) FN increase, (d) TPR, (e) Precision, (f) F1 measure. } %
\label{fig:soldo}
\vspace{0.15cm}
\end{figure*}

\section{Evaluating CPB Techniques}

\subsection{Soldo et al.~\cite{soldo2010predictive}}
\label{sec:soldo_approach}
We first evaluate Soldo et al~\cite{soldo2010predictive}'s CPB approach based on implicit recommendation.
We do so aiming to: (1) evaluate false positives and false negatives,
which were not taken into consideration in~\cite{soldo2010predictive}, and (2) compare against privacy-friendly approaches, presented later.
Essentially, Soldo et al.'s work builds on~\cite{zhang2008highly}, which bases on a relevance ranking scheme similar to PageRank, measuring the correlation of an attacker to a contributor relying on their history as well as the attacker's recent log production patterns. Soldo et al.~significantly improve on this, by using an implicit recommendation system to discover similar victims as well as groups of correlated victims and attackers. The presence of attacks performed by the same source around the same time leads to stronger victim similarity, and a neighborhood model (k-NN) is applied to cluster similar victims.
Cross Association (CA) co-clustering~\cite{chakrabarti2004fully} is then used to discover groups of correlated attackers and victims, and prediction within the cluster is done via the EWMA time series algorithm (TS) to capture attacks' temporal trends.
\edit{In other words, the prediction score for each organization is a weighted ensemble of three methods (TS, k-NN and CA).}
We have re-implemented their system in Python, using Chakrabarti's CA implementation~\cite{chakrabarti2004fully}. %

We start by measuring the basic predictor which only relies on a local EWMA time series algorithm (TS), using $\alpha=0.9$ as it \edit{yields the best results}, then, apply the co-clustering techniques (TS-CA), and, finally, implement their full scheme by combining k-NN to cluster victims based on their similarity with CA and TS (TS-CA-k-NN).
Fig.~\ref{fig:soldo} illustrates the improvement/increase in TP, FP, FN (compared to the TS baseline)
as well as TPR, PPV, and F1,
with various $k$ values (ranging from 1 to 35) used by the k-NN algorithm to discover similar organizations. Obviously, the k-NN parameter $k$ does not affect TS-CA and TS.  

Fig.~\ref{fig:soldo-tp-incr} shows that, with TS-CA-k-NN, $TP_{impr}$ increases significantly with $k$, 
almost doubling the ``hit count'' compared to the TS baseline, whereas, TS-CA improves less ($0.67$).
On the other hand, however, there is $FP_{incr}$ too, 5- to 50-fold, as clusters become bigger (Fig.~\ref{fig:soldo-fp-incr}), and naturally, this stark increase in FP leads to low precision, as shown in Fig.~\ref{fig:soldo-ppv}.
FNs also always increase compared to TS (Fig.~\ref{fig:soldo-fn-incr}), specifically, they double
with TS-CA and increase between $0.55$ and $0.99$ (less for larger $k$ values) compared to TS. 
$FN_{incr}$ also affects TPR (Fig.~\ref{fig:soldo-tpr}), with an increase between $0.58$ and $0.66$.
The $TP_{impr}$ does not correspond to a comparable increase in TPR, due to the poor FN performance, as shown by the fact that TS-CA-k-NN reaches $0.99$ in $TP_{impr}$ but only at most $0.66$ TPR compared to $0.59$ with the baseline TS.
Overall, Soldo et al.'s techniques achieve poor F1 measures, at most $0.16$ and $0.14$, with TS-CA and TS-CA-k-NN, actually lower than a simple local time-series prediction ($0.26$).

\begin{figure*}[t]
\centering
\begin{subfigure}[t]{0.325\textwidth}
\centering
\includegraphics[width=1\textwidth,height=0.18\textheight]{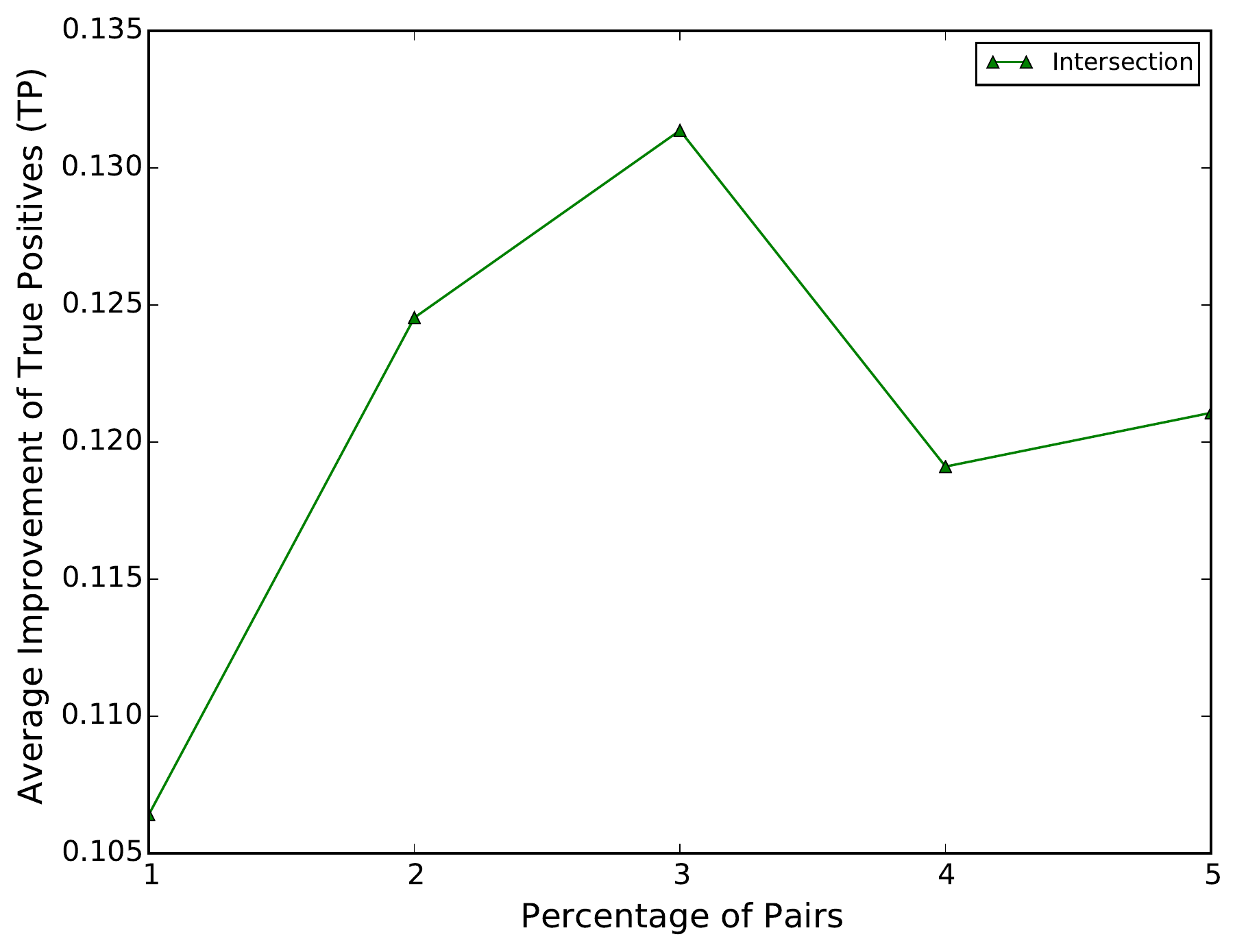}
\caption{ %
\label{fig:dimva-global-tp-impr}}
\end{subfigure}
\begin{subfigure}[t]{0.325\textwidth}
\centering
\includegraphics[width=1\textwidth,height=0.18\textheight]{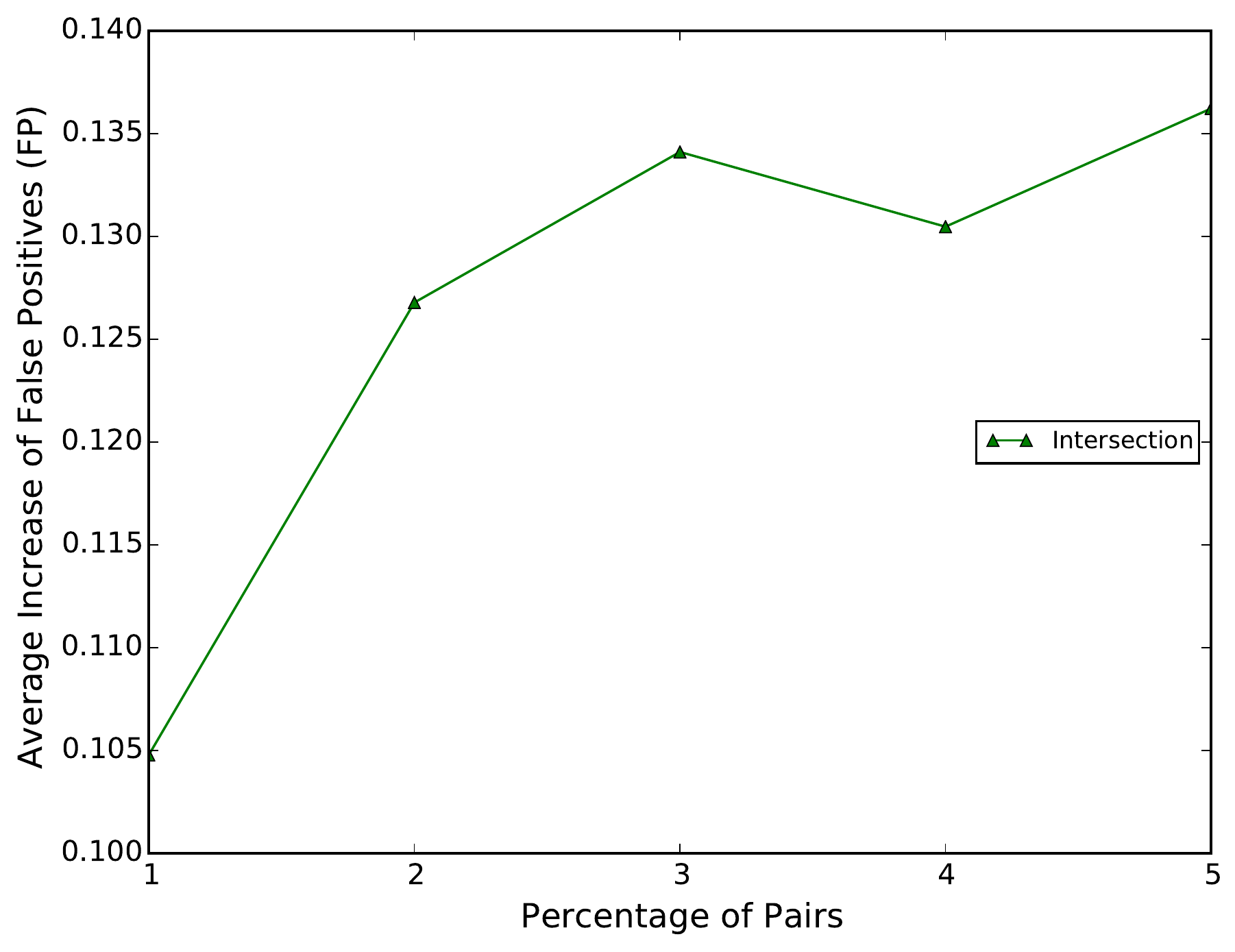}
\caption{ %
\label{fig:dimva-global-fp-impr}}
\end{subfigure}
\begin{subfigure}[t]{0.325\textwidth}
\centering
\includegraphics[width=1\textwidth,height=0.18\textheight]{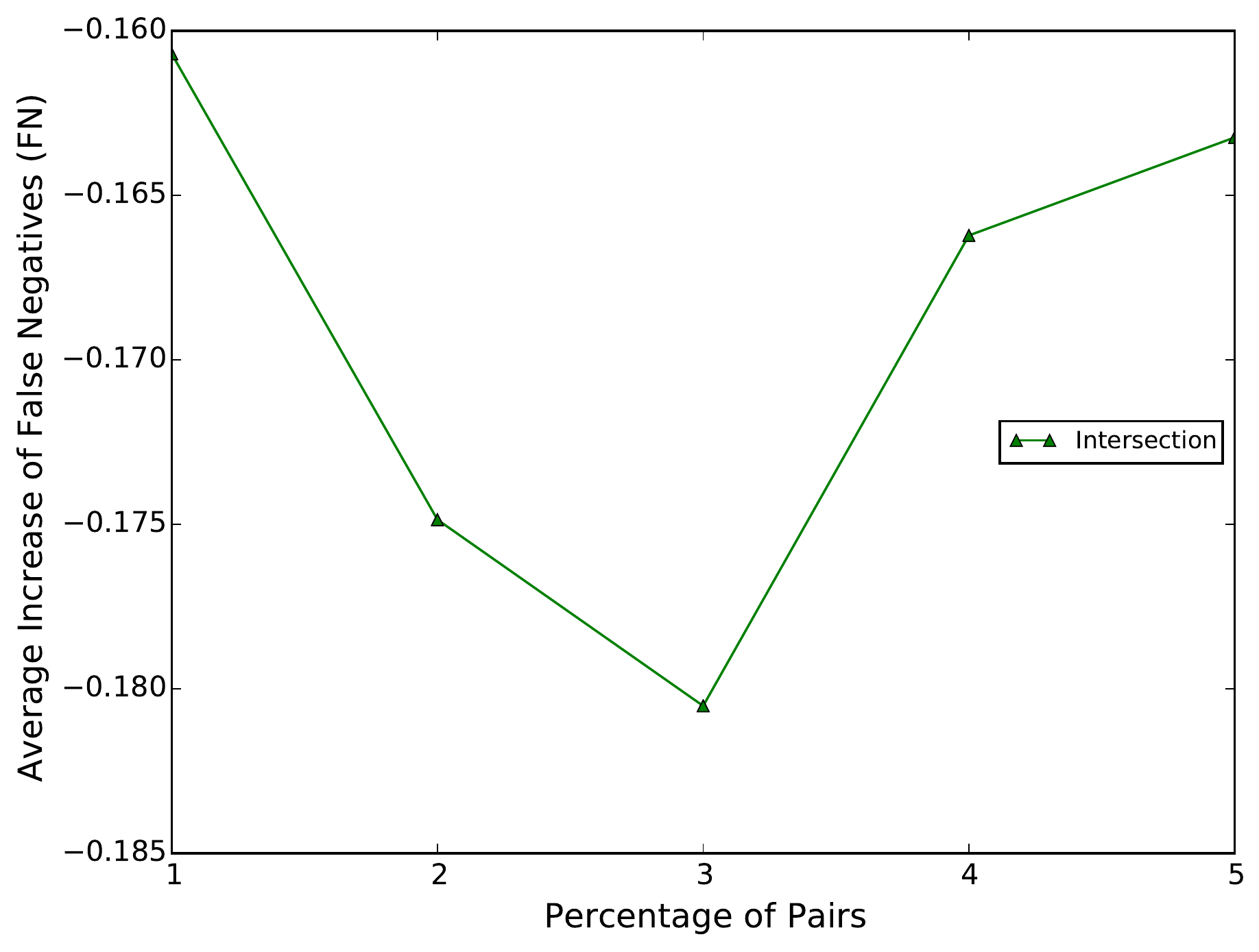}
\caption{ %
\label{fig:dimva-global-fn-impr}}
\end{subfigure}
\begin{subfigure}[t]{0.325\textwidth}
\centering
\includegraphics[width=1\textwidth,height=0.18\textheight]{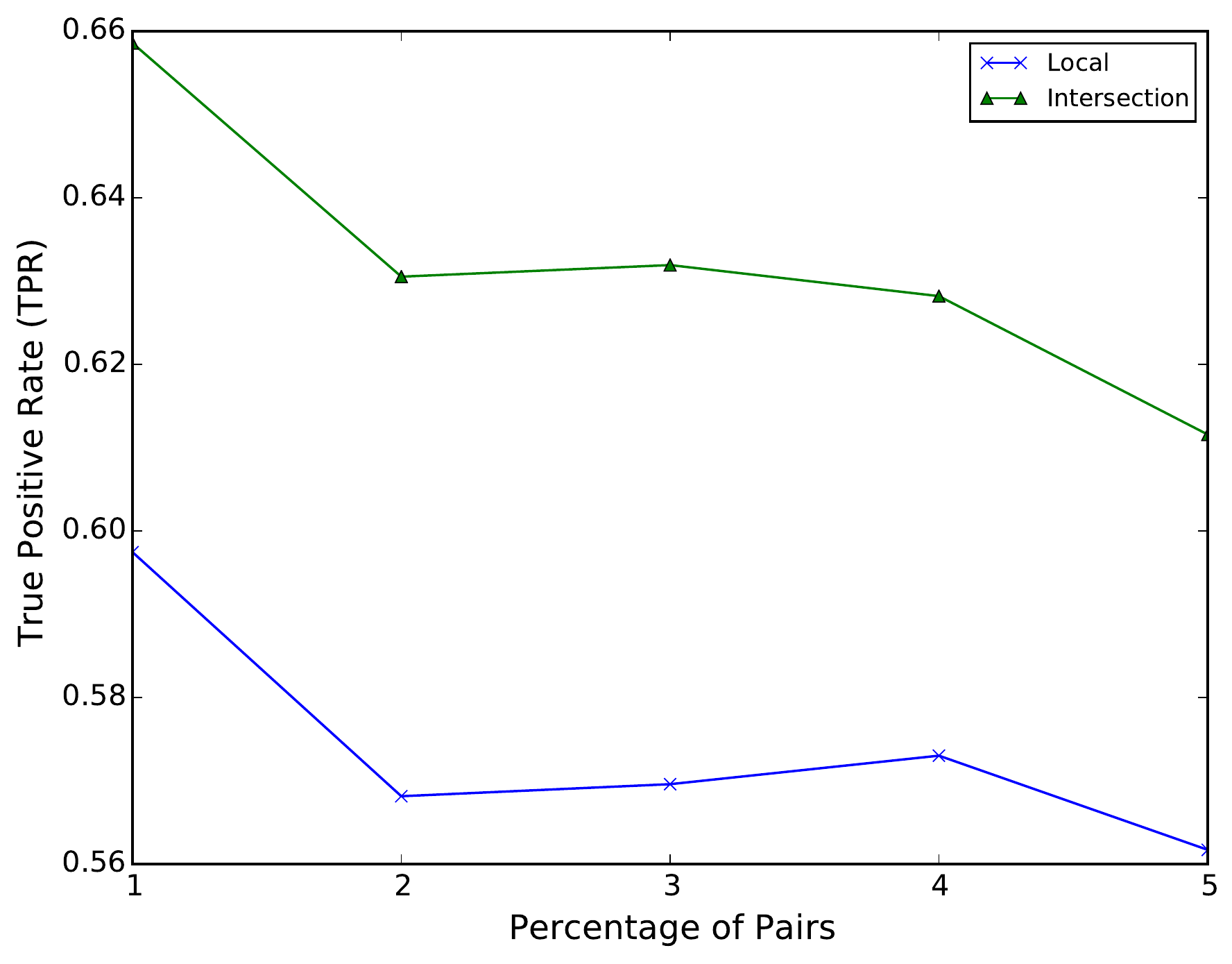}
\caption{ %
\label{fig:dimva-global-TPR}}
\end{subfigure} 
\begin{subfigure}[t]{0.325\textwidth}
\centering
\includegraphics[width=1\textwidth,height=0.18\textheight]{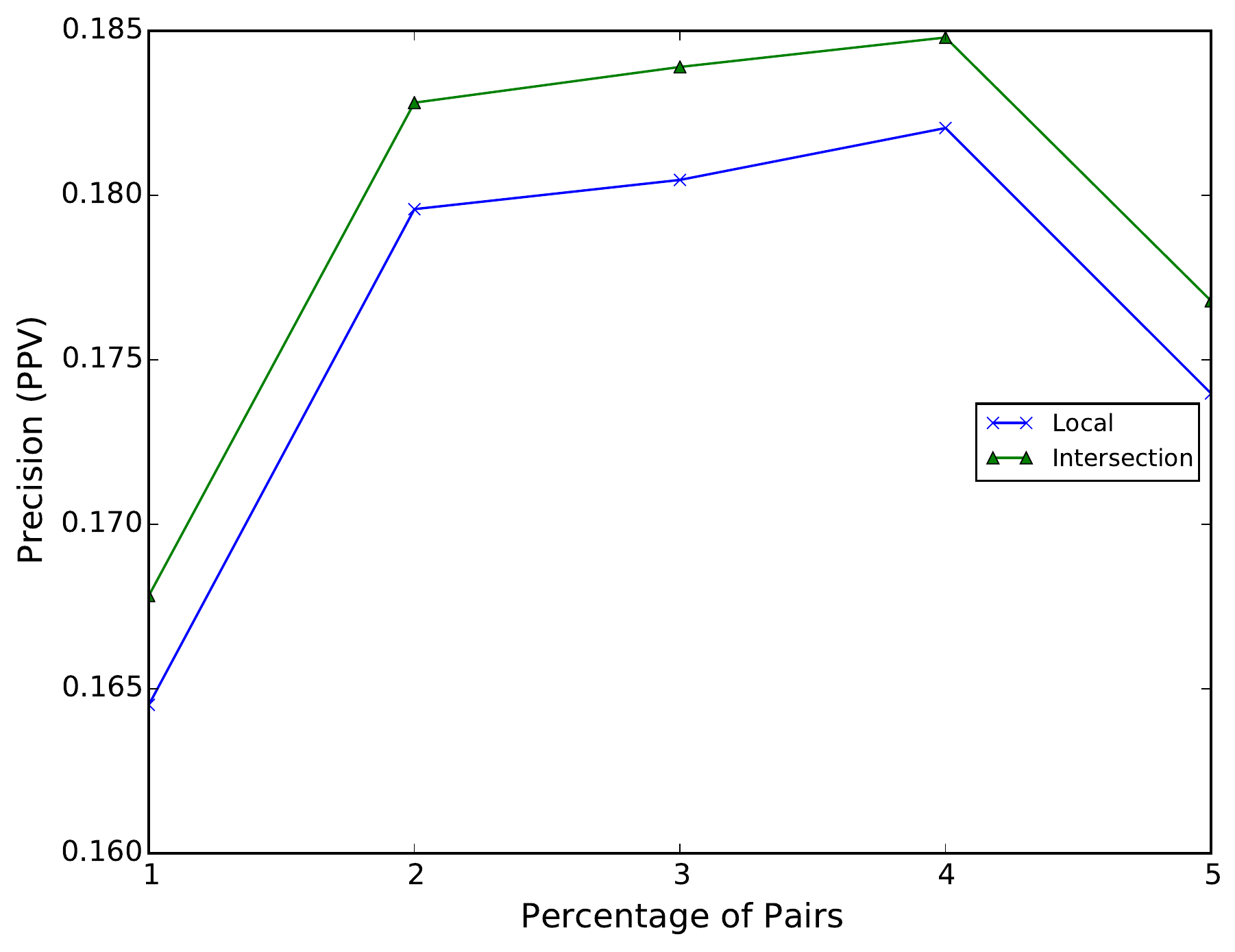}
\caption{ %
\label{fig:dimva-global-PPV}}
\end{subfigure}
\begin{subfigure}[t]{0.325\textwidth}
\centering
\includegraphics[width=1\textwidth,height=0.18\textheight]{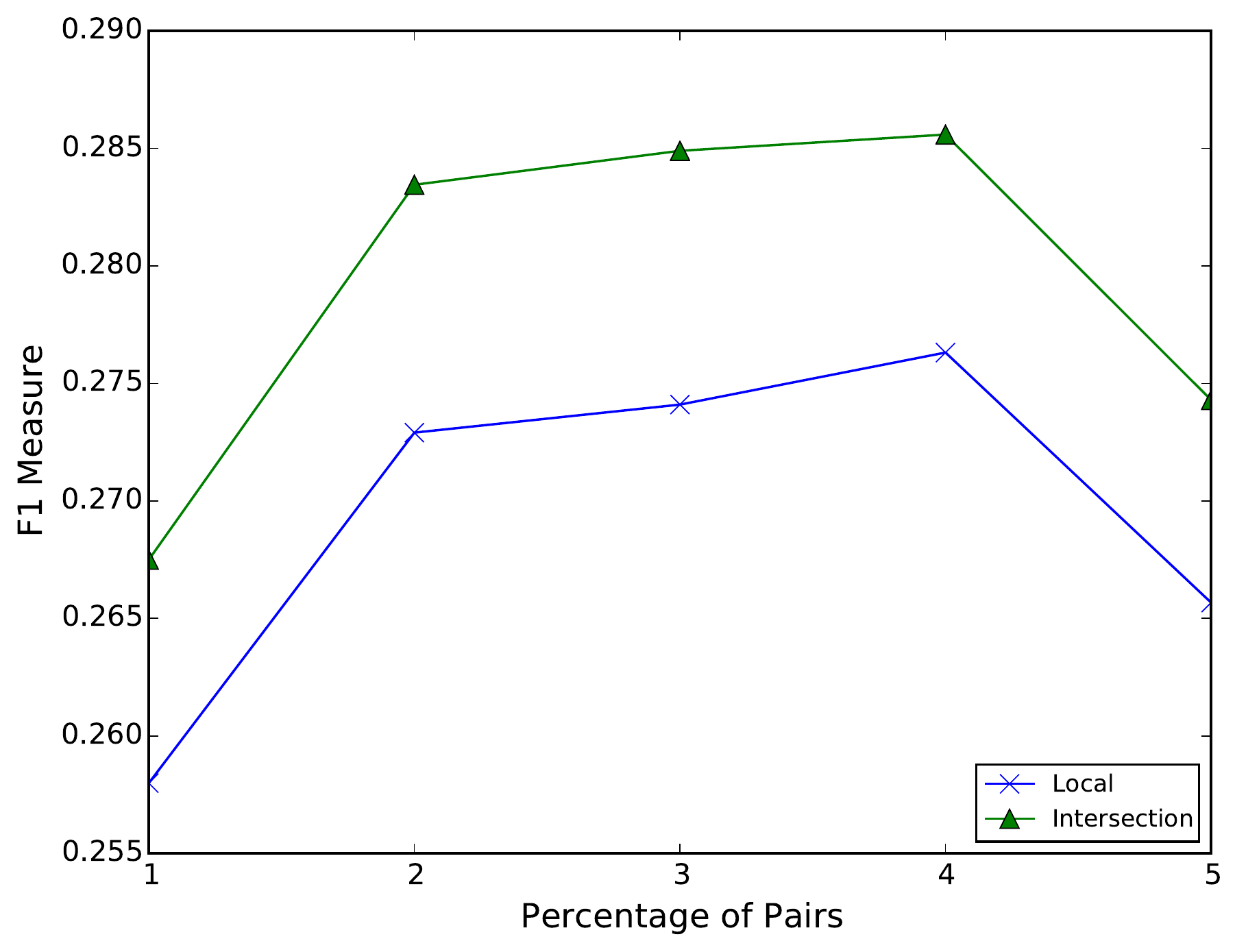}
\caption{ %
\label{fig:dimva-global-f1}}
\end{subfigure} 
\vspace{-0.15cm}
\caption{ Freudiger et al.~\cite{freudiger2015controlled} (a) TP improvement, (b) FP increase, (c) FN increase, (d) TPR, (e) Precision and (f) F1, with increasing percentage of global pairs.}
\label{fig:dimva-global}
\vspace{-0.15cm}

\end{figure*}

\subsection{Freudiger et al.~\cite{freudiger2015controlled}}\label{sec:freudiger}
Next, we evaluate the privacy-friendly peer-to-peer approach to CPB by Freudiger et al.~\cite{freudiger2015controlled}. Organizations interact pairwise, aiming to privately estimate the benefits of collaboration,
and then share data with entities that are likely to yield the most benefits.
They also use DShield data and perform prediction using EWMA.
They find that: (1) the number of common attacks is the best predictor of benefits, which can be estimated
privately, using Private Set Intersection Cardinality (PSI-CA)~\cite{de2012fast}; and 
(2) sharing only the intersection of attacks -- which can be done privately using Private Set Intersection (PSI)~\cite{de2010practical} --
is almost as beneficial as sharing everything.
Their goal is really to assess benefit estimation/sharing strategies, rather than to focus on deployment. They assume a network of 100 organizations, select the ``top 50'' among all possible 4950 pairs (in terms of estimated benefits), and only experiment on those. Naturally, without a coordinating entity, it is impossible to rank the pairs,
so they suggest that one should collaborate with either organizations when estimated benefits are above a threshold, although it is not stated how to set this threshold; or with the top $x$ organizations with the biggest estimated benefits, but do not experiment with or discuss how $x$ impacts overhead or true/false positives. 
We replicate both approaches: {\bf (A)} with the top 1\% to 5\% of global pairs, and {\bf (B)} having each organization pick $1$ to $35$ most similar organizations. %

Fig.~\ref{fig:dimva-global} shows the improvement/increase in TP, FP, FN (compared to a baseline with no sharing) as well as TPR, PPV and F1 with increasing percentage of global pairs (A). 
We omit plots for approach (B) since they are worse across the board, although we discuss them next.
Looking at $TP_{impr}$, (A) %
yields $13\%$ increase when $3\%$ of global pairs are selected whereas for (B), i.e. picking local pairs, $TP_{impr}$ increases along with the number of local pairs selected. 
(A) has a rather small $FP_{incr}$ ($13\%$ increase when the $75$ top pairs are selected) compared to (B) which is affected by the number of pairs that each organizations picks for collaboration. When an organization collaborates with 5 others a $25\%$ $FP_{incr}$ is observed on average while when it collaborates with 30 others $FP_{incr}$ reaches $80\%$. Moreover, 
we find both approaches achieve a decrease in false negatives with the second approach achieving bigger decreases as the number of collaborators increases. 

Overall, both approaches improve precision and recall of the system, yielding higher F1 scores compared to a local approach. Although the increase in TP is not as high as with the non-private approach of~\cite{soldo2010predictive}, a more balanced increase of false positives and a decrease of the false negatives seems possible.
However, the system is limited in the amount of new information organizations learn (e.g., only events about IPs they have already seen is shared) as well as scalability, \edit{since both the computation of the metrics and the actual data sharing are conducted pair-wise (if there are $n$ collaborating entities, the complexity of the data sharing would be $O(n^2)$).}

\section{A Novel Hybrid Approach}
\label{section:methodology}
 
\edit{Centralized state-of-the-art CPB techniques~\cite{soldo2010predictive} have only focused on improving ``hit counts,'' but, as shown above, they generate very high false positive rates. 
In practice, organizations might not adopt such solutions if they generate a large number of false alarms.
Naturally, one could design better centralized approaches that yield better accuracy, e.g., by learning to discard the data that yield false positives.
However, our intuition is that in this case a privacy-preserving approach might be best suited as it can (i) help organizations find a better trade-off between prediction improvement and increase in false positives, and (ii) do so while actually minimizing exposure of possibly confidential data.}

\descr{Overview.} To this end, we introduce a novel hybrid  system which relies on a semi-trusted authority, or \STA, acting as a coordinating entity to facilitate clustering without having access to the raw data. In other words, the \STA clusters contributors based on the similarity of their logs (without accessing these logs), and helps organizations in the same cluster to share relevant logs. 

The system involves four steps. (1) First, organizations interact in a pairwise manner to privately compute
a similarity measure of their logs, based on the number of common attacks (similar to~\cite{freudiger2015controlled}).
Then, (2) the \STA collects the similarity measures from each organization and performs clustering using
one of three possible algorithms, i.e., Agglomerative Clustering, k-means, or k-NN.\footnote{Note that, to ease presentation, we do not plot results using Agglomerative Clustering because it yields the worse results.}
Next, (3) the \STA reports to each organization the identifiers of other organizations in the same cluster (if any), so that they collaboratively, yet privately, share logs to boost the accuracy of their prediction, by either sharing common attacks ({\em intersection}), correlated attacks ({\em \iptip}), or both. 
For comparison, we also consider baseline approaches, i.e.,
sharing nothing ({\em local}) or sharing everything ({\em global}).
Finally, (4) each organization performs EWMA prediction (again, with $\alpha=0.9$, as done in our evaluation of~\cite{soldo2010predictive}).
based on their logs, plus those from entities in the same cluster. 
This approach is {\em hybrid} in that, while involving a central authority, data sharing is privacy-friendly: in (1) the number of common attacks can be computed using PSI-CA~\cite{de2012fast}, while in (3) sharing of common attacks can occur using PSI~\cite{de2010practical} and of correlated attacks using~\cite{melis}.

\begin{figure*}[!t]
\begin{subfigure}[t]{0.325\textwidth}
\centering
\includegraphics[width=1\textwidth,height=0.18\textheight]{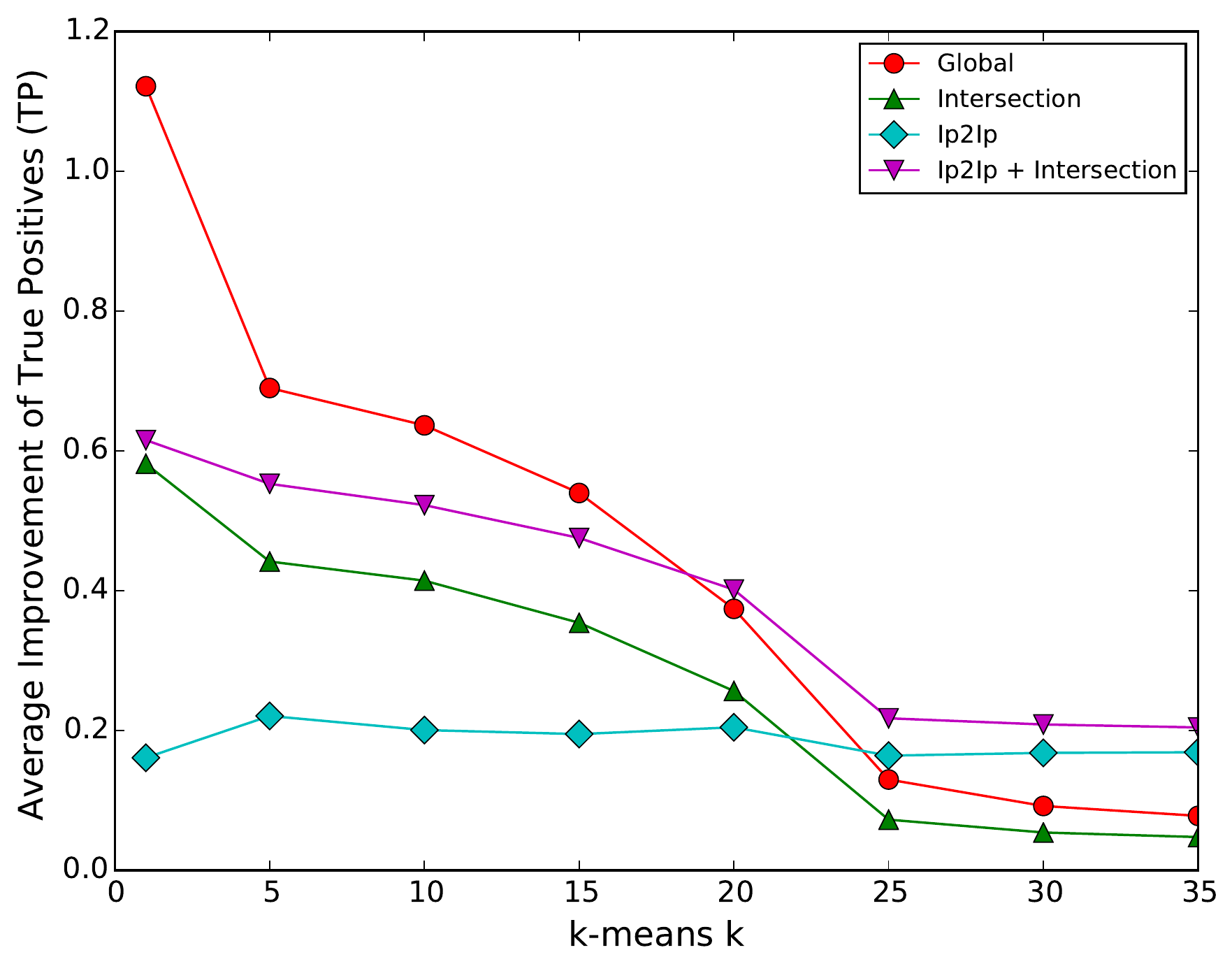}
\caption{\label{fig:kmeans-tp-incr}}
\end{subfigure} 
\begin{subfigure}[t]{0.325\textwidth}
\centering
\includegraphics[width=1\textwidth,height=0.18\textheight]{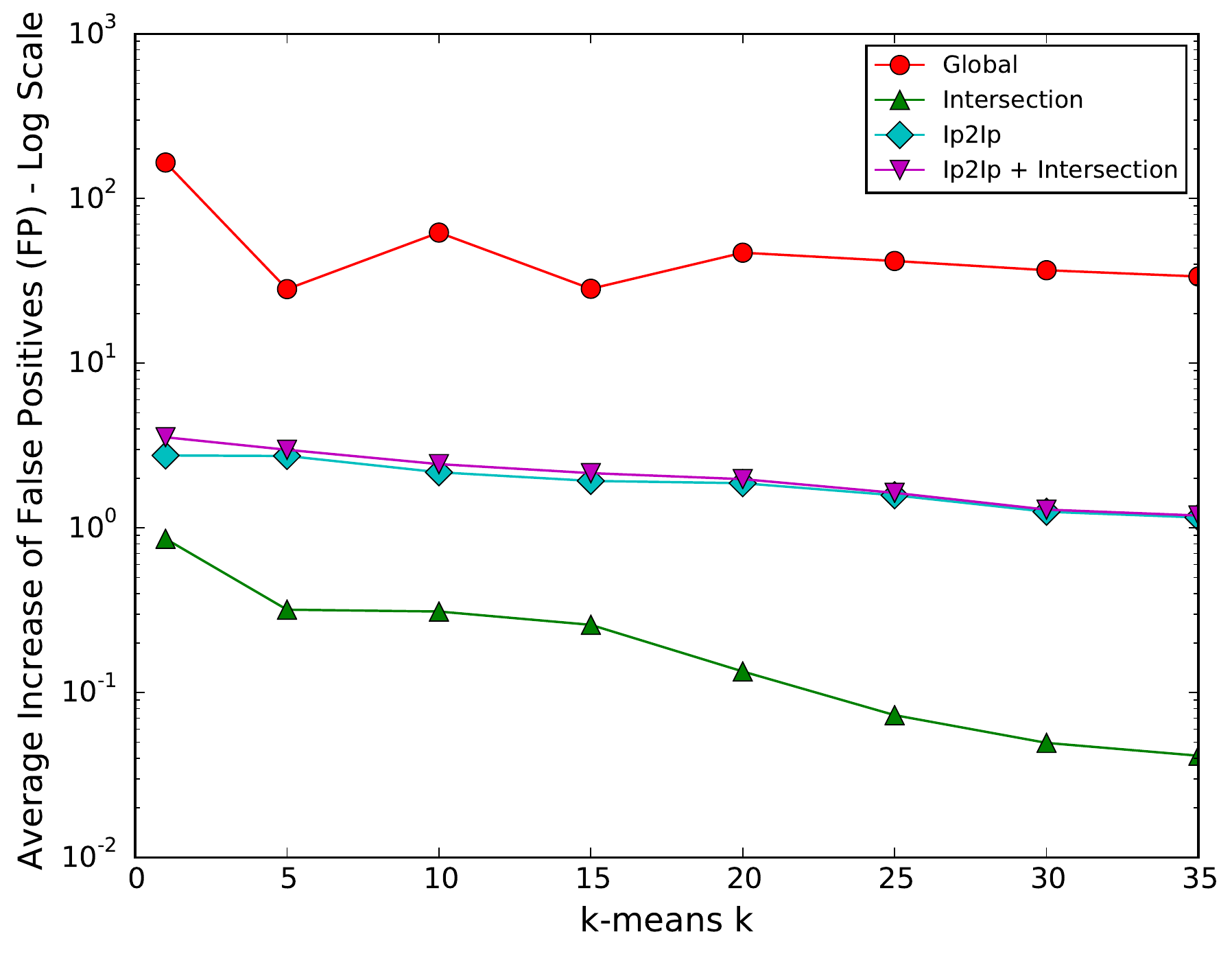}
\caption{\label{fig:kmeans-fp-incr}}
\end{subfigure} 
\begin{subfigure}[t]{0.325\textwidth}
\centering
\includegraphics[width=1\textwidth,height=0.18\textheight]{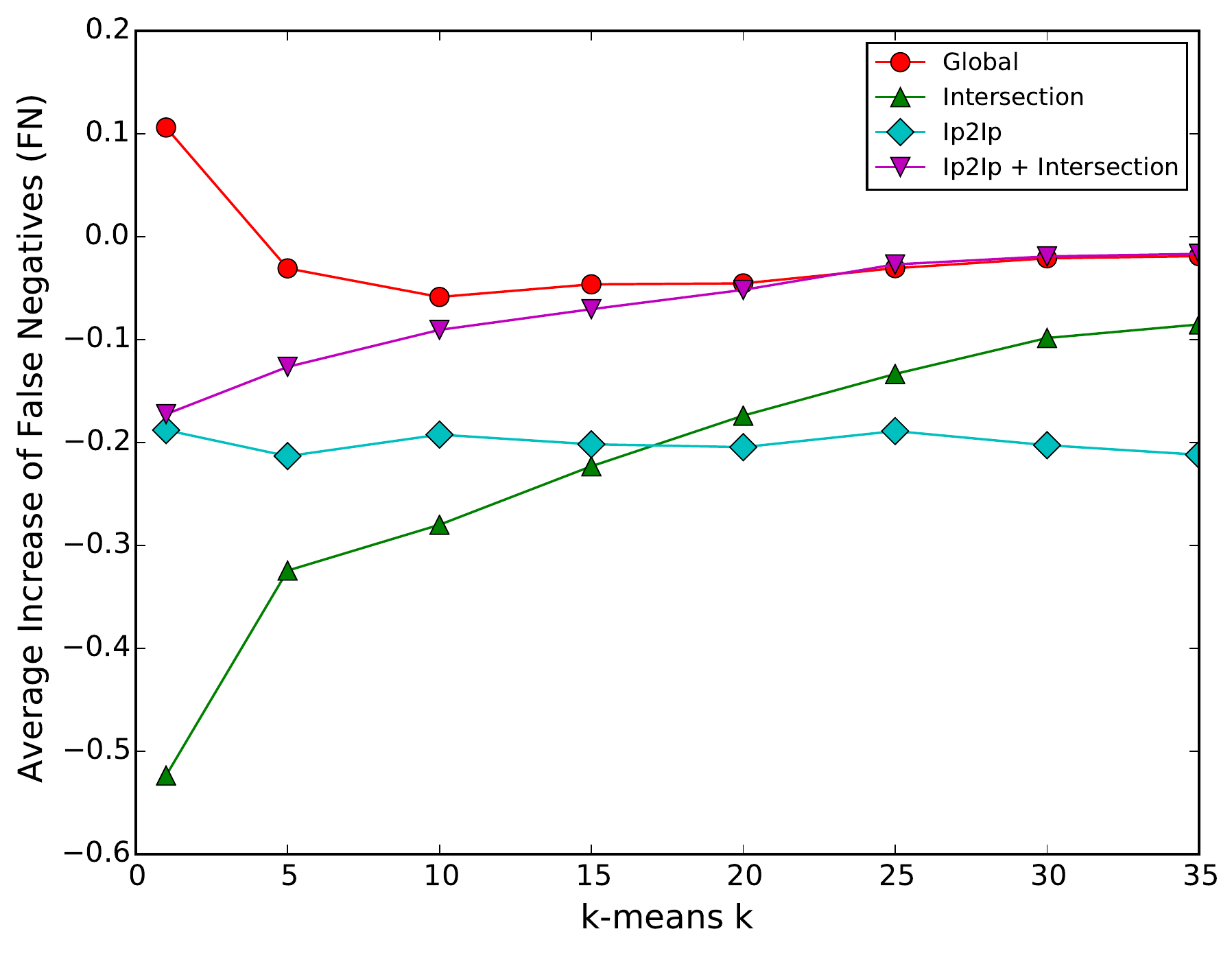}
\caption{\label{fig:kmeans-fn-incr}}
\end{subfigure}
\centering
\begin{subfigure}[t]{0.325\textwidth}
\centering
\includegraphics[width=1\textwidth,height=0.18\textheight]{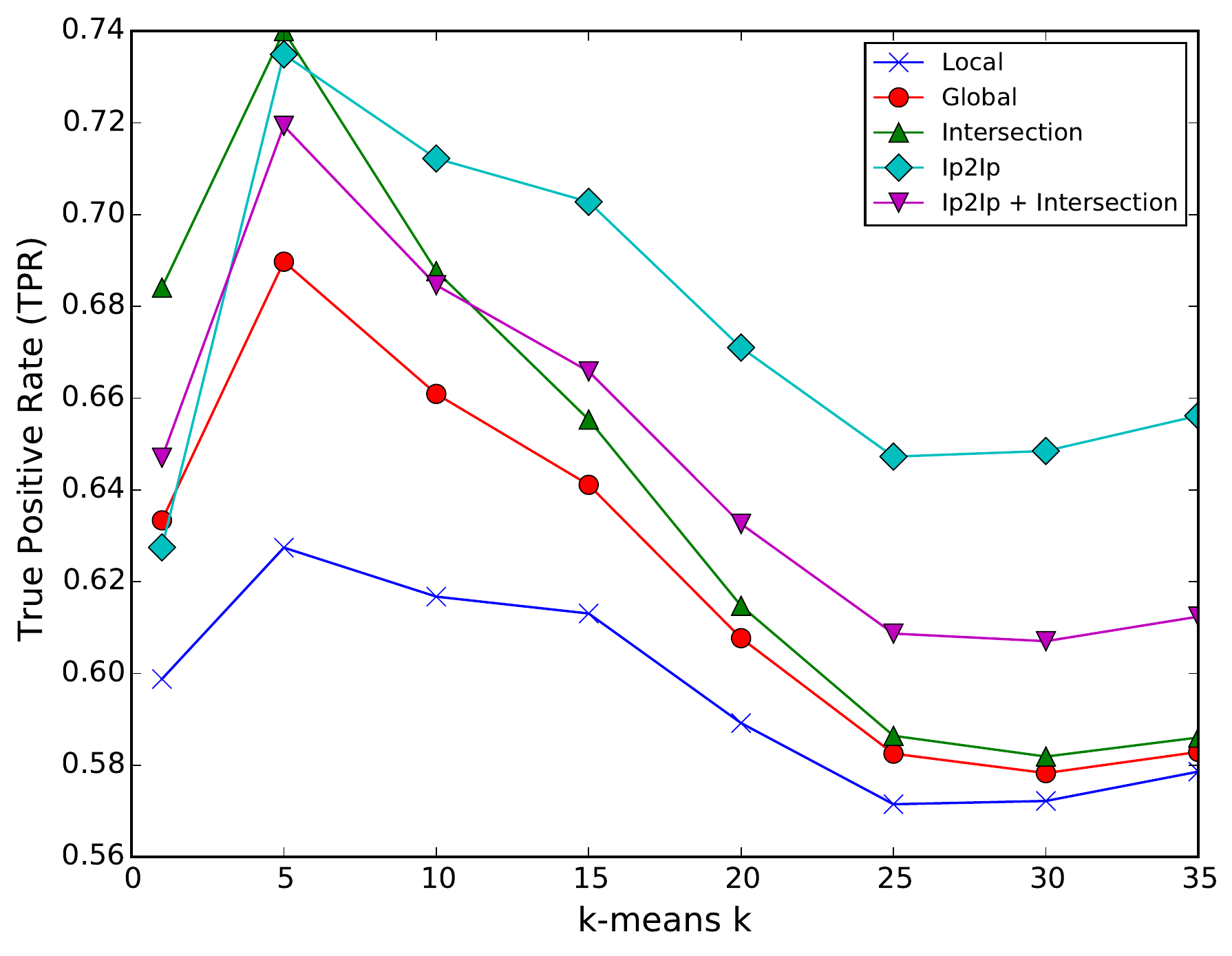}
\caption{\label{fig:kmeans-tpr}}
\end{subfigure} 
\begin{subfigure}[t]{0.325\textwidth}
\centering
\includegraphics[width=1\textwidth,height=0.18\textheight]{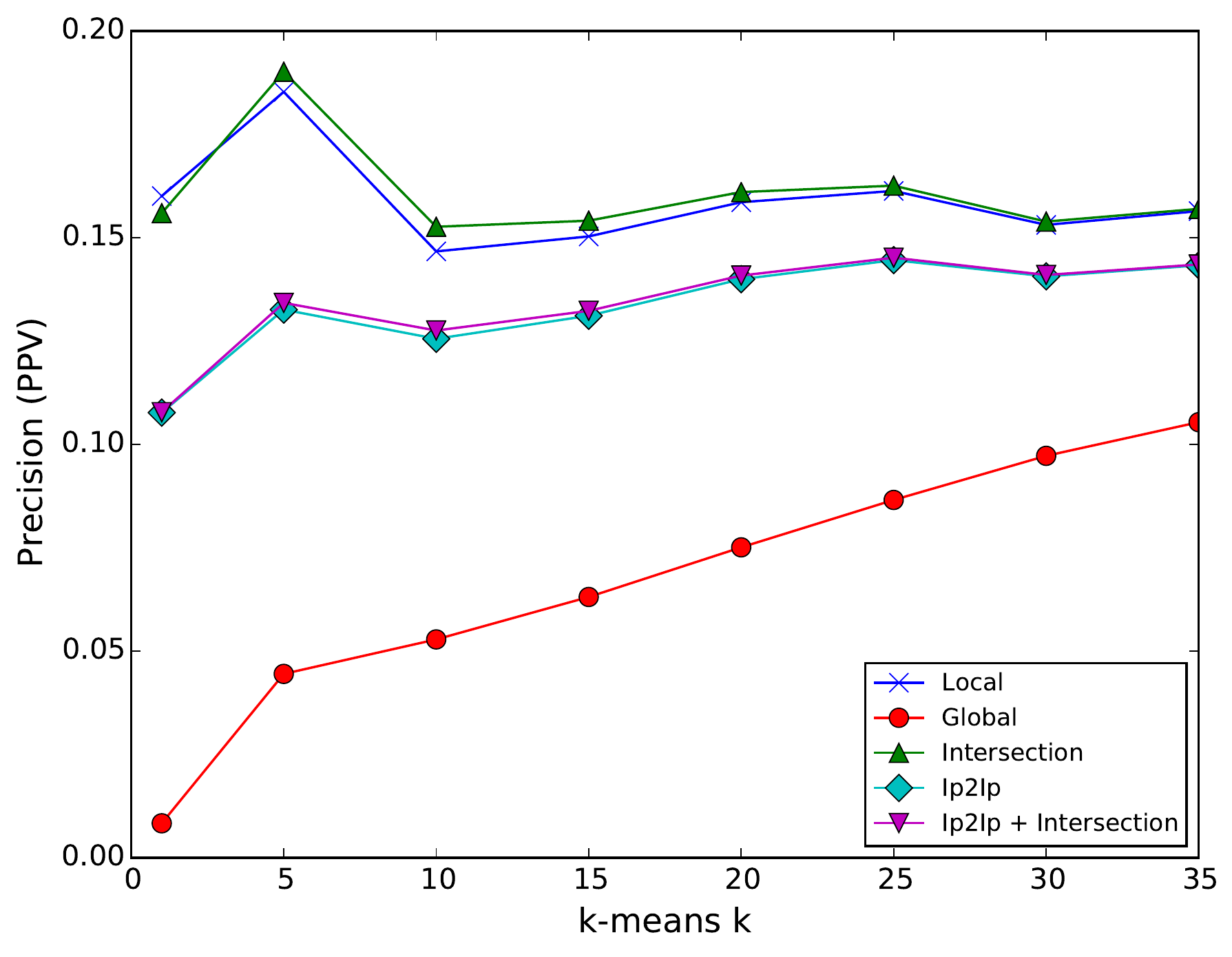}
\caption{\label{fig:kmeans-ppv}}
\end{subfigure} 
\begin{subfigure}[t]{0.325\textwidth}
\centering
\includegraphics[width=1\textwidth,height=0.18\textheight]{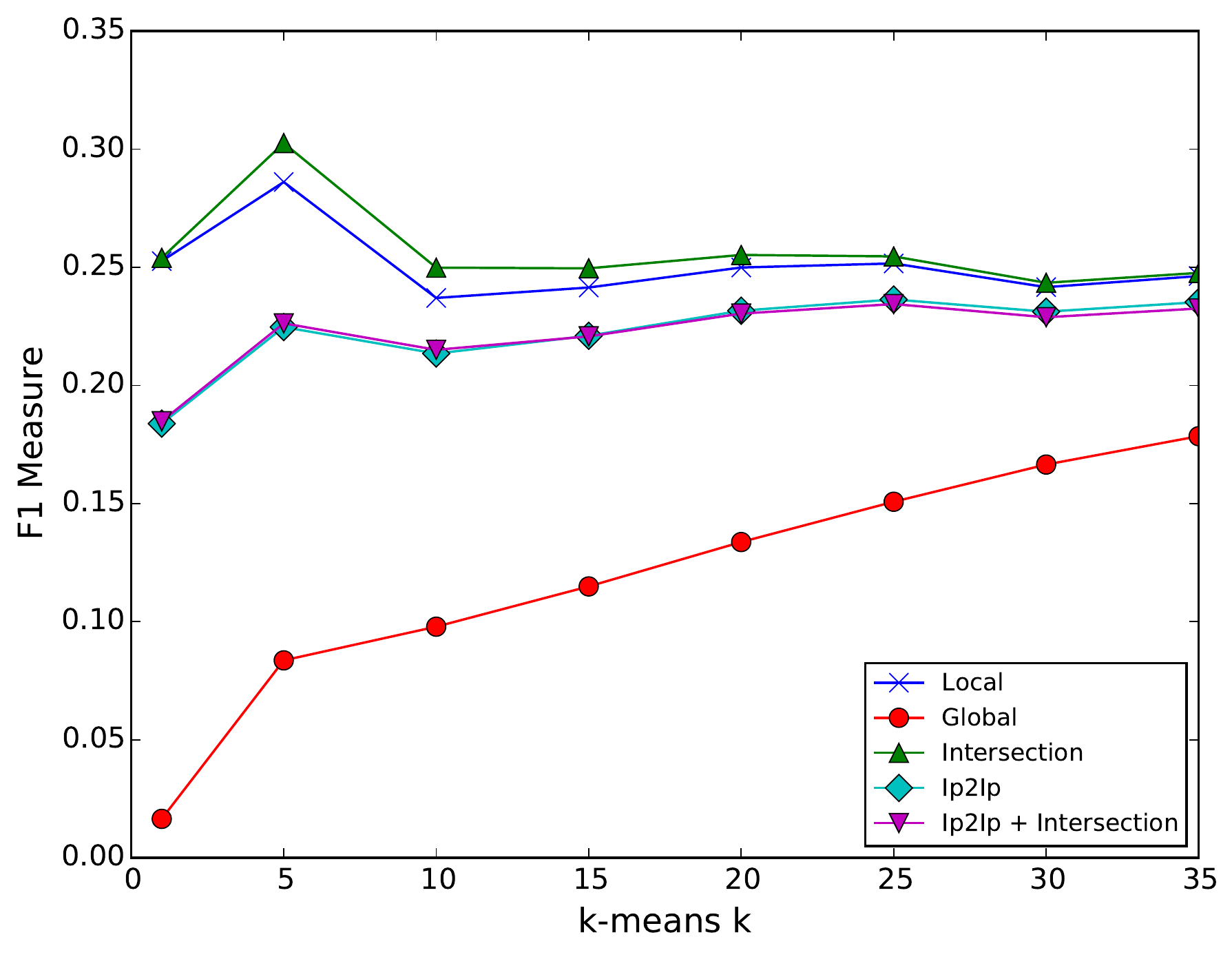}
\caption{\label{fig:kmeans-f1}}
\end{subfigure}  \\ 
\vspace{-0.35cm}
\caption{k-means: (a) TP improvement, (b) FP increase (y-axis in log scale), (c) FN increase, (d) TPR, (e) Precision, (f) F1 measure.} %
\vspace{-0.35cm}
\label{fig:kmeans}
\end{figure*}

\begin{figure*}[!t]
\centering
\begin{subfigure}[t]{0.325\textwidth}
\centering
\includegraphics[width=1\textwidth,height=0.18\textheight]{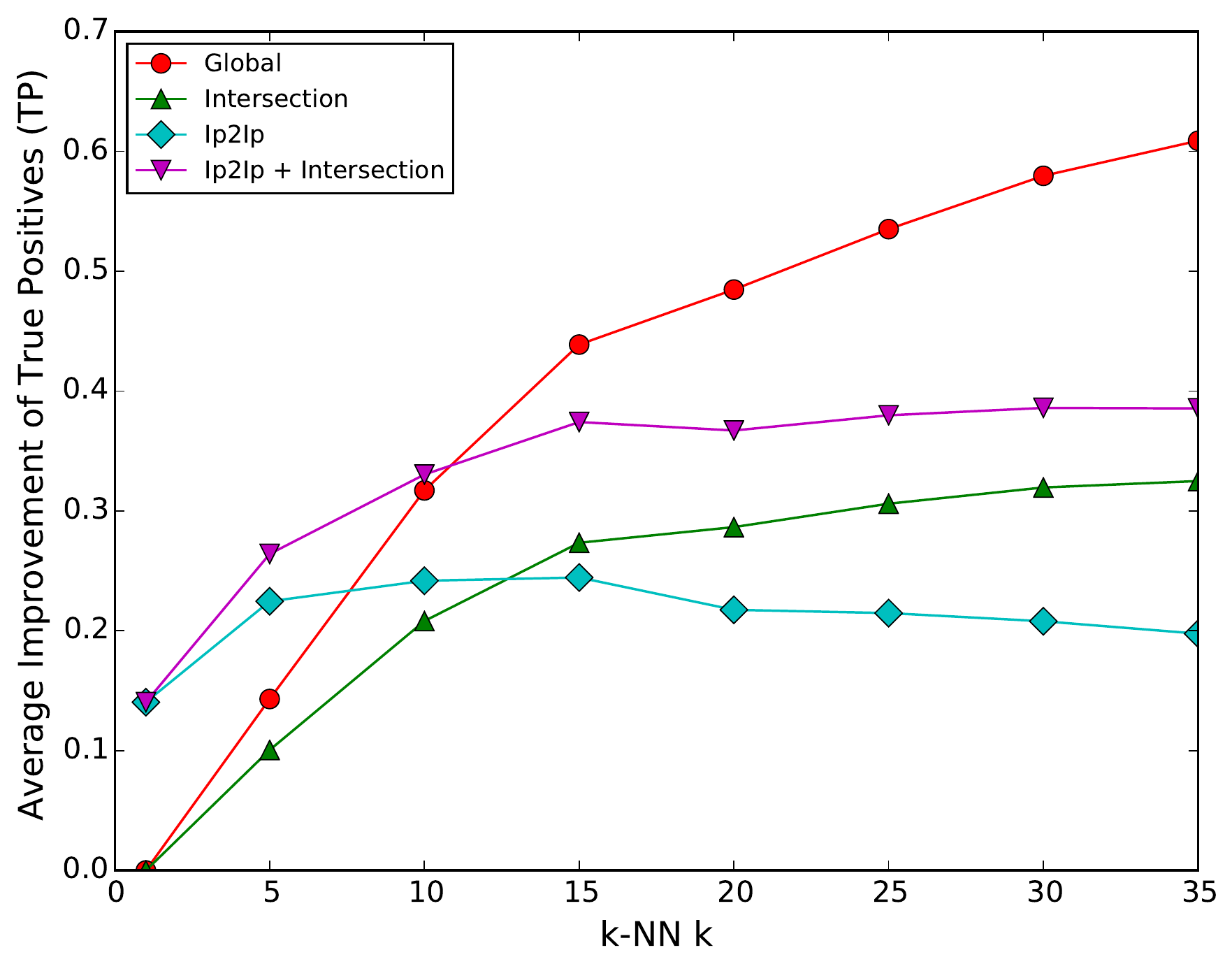}
\caption{\label{fig:knn-tp-increase}}
\end{subfigure}
\begin{subfigure}[t]{0.325\textwidth}
\centering
\includegraphics[width=1\textwidth,height=0.18\textheight]{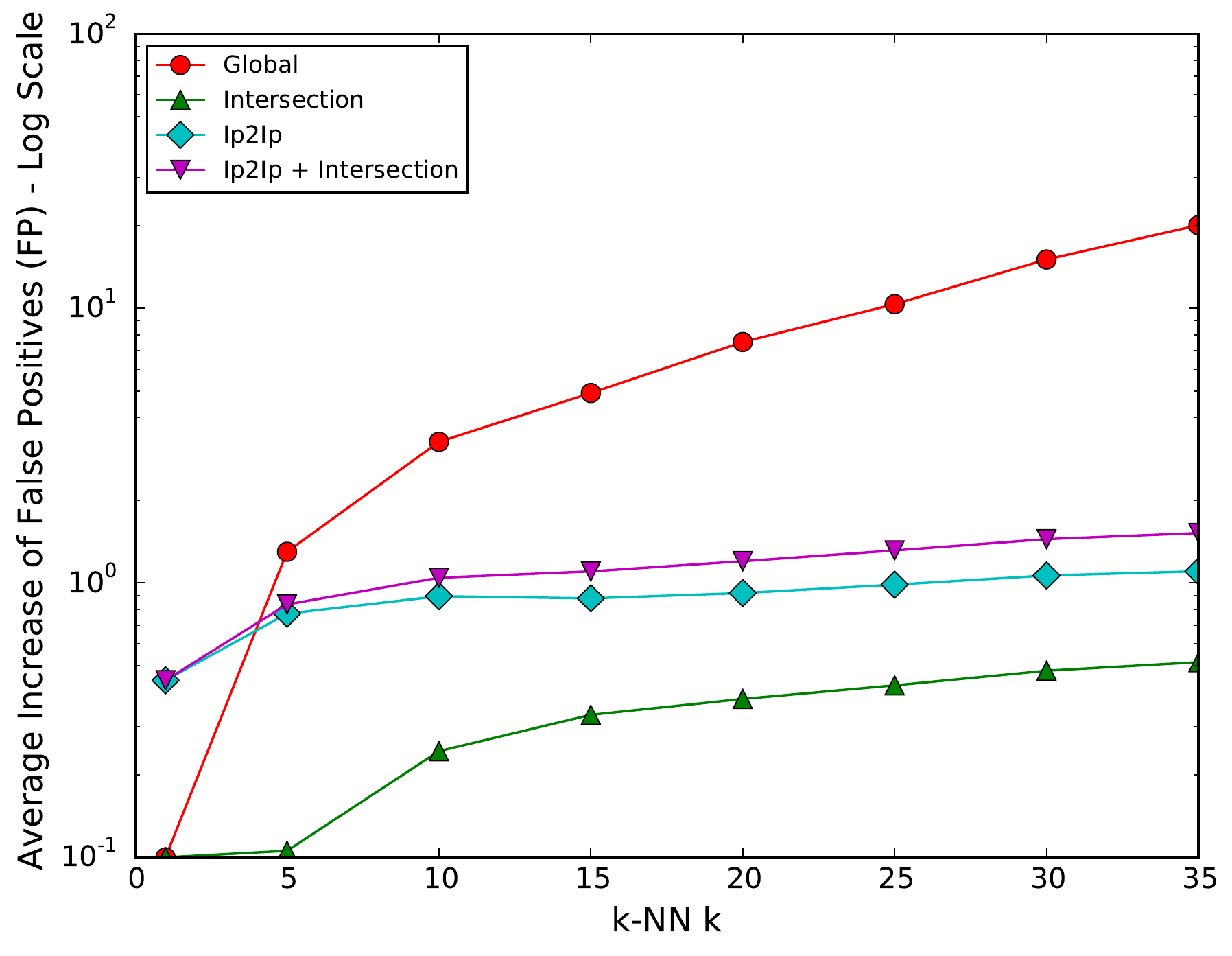}
\caption{\label{fig:knn-fp-increase}}
\end{subfigure}
\begin{subfigure}[t]{0.325\textwidth}
\centering
\includegraphics[width=1\textwidth,height=0.18\textheight]{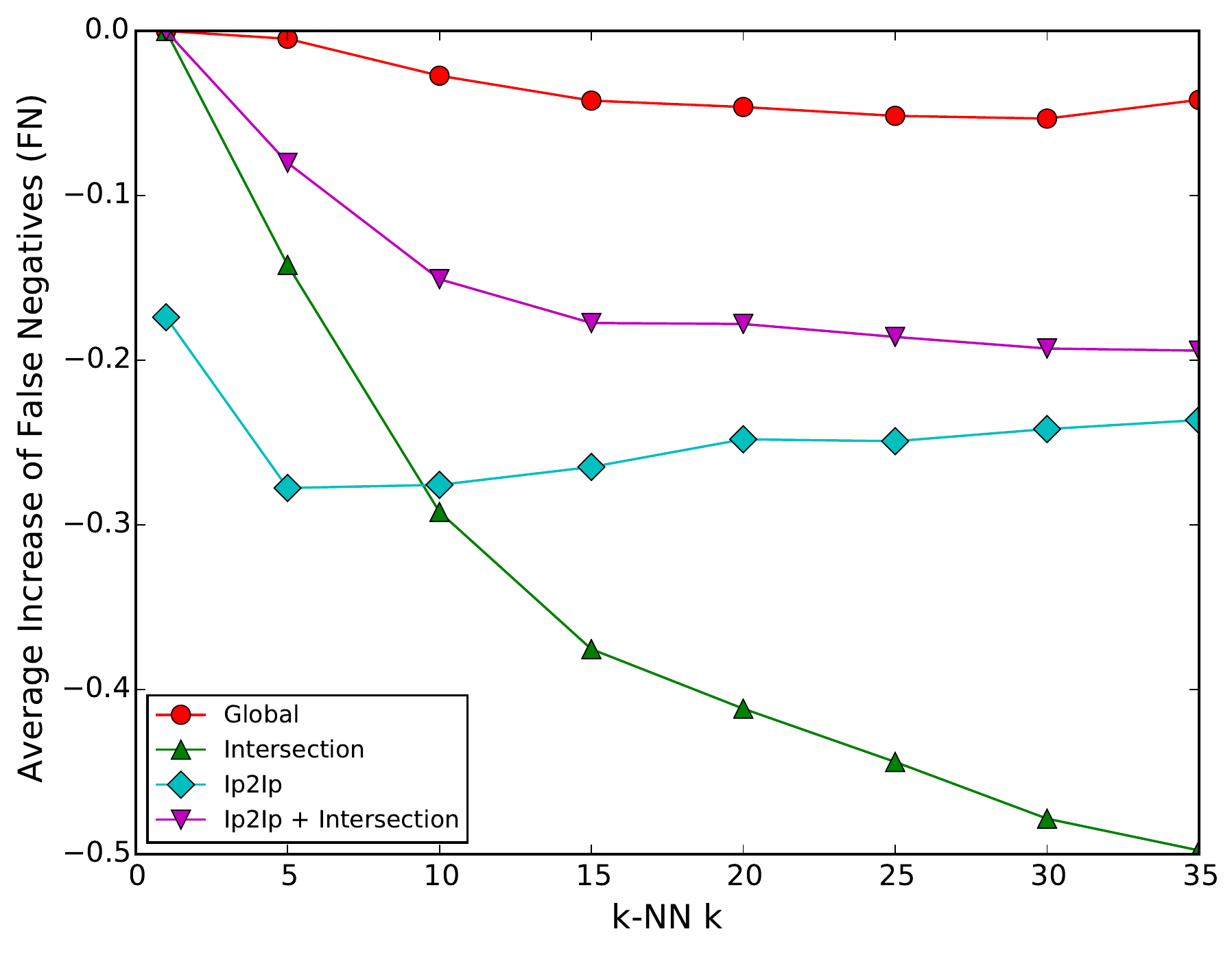}
\caption{\label{fig:knn-fn-incr}}
\end{subfigure}
\\
\begin{subfigure}[t]{0.325\textwidth}
\centering
\includegraphics[width=1\textwidth,height=0.18\textheight]{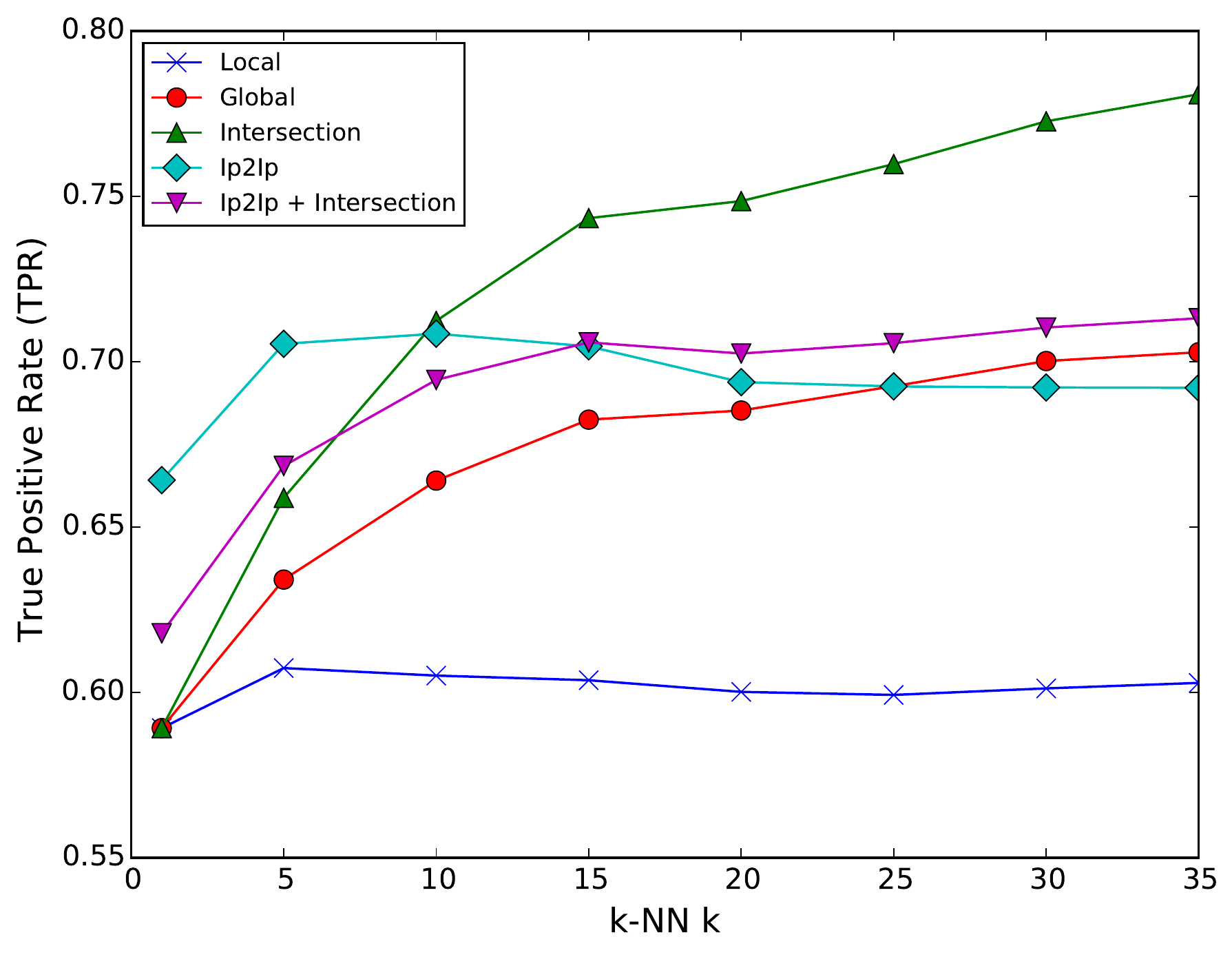}
\caption{\label{fig:knn-tpr}}
\end{subfigure}
\begin{subfigure}[t]{0.325\textwidth}
\centering
\includegraphics[width=1\textwidth,height=0.18\textheight]{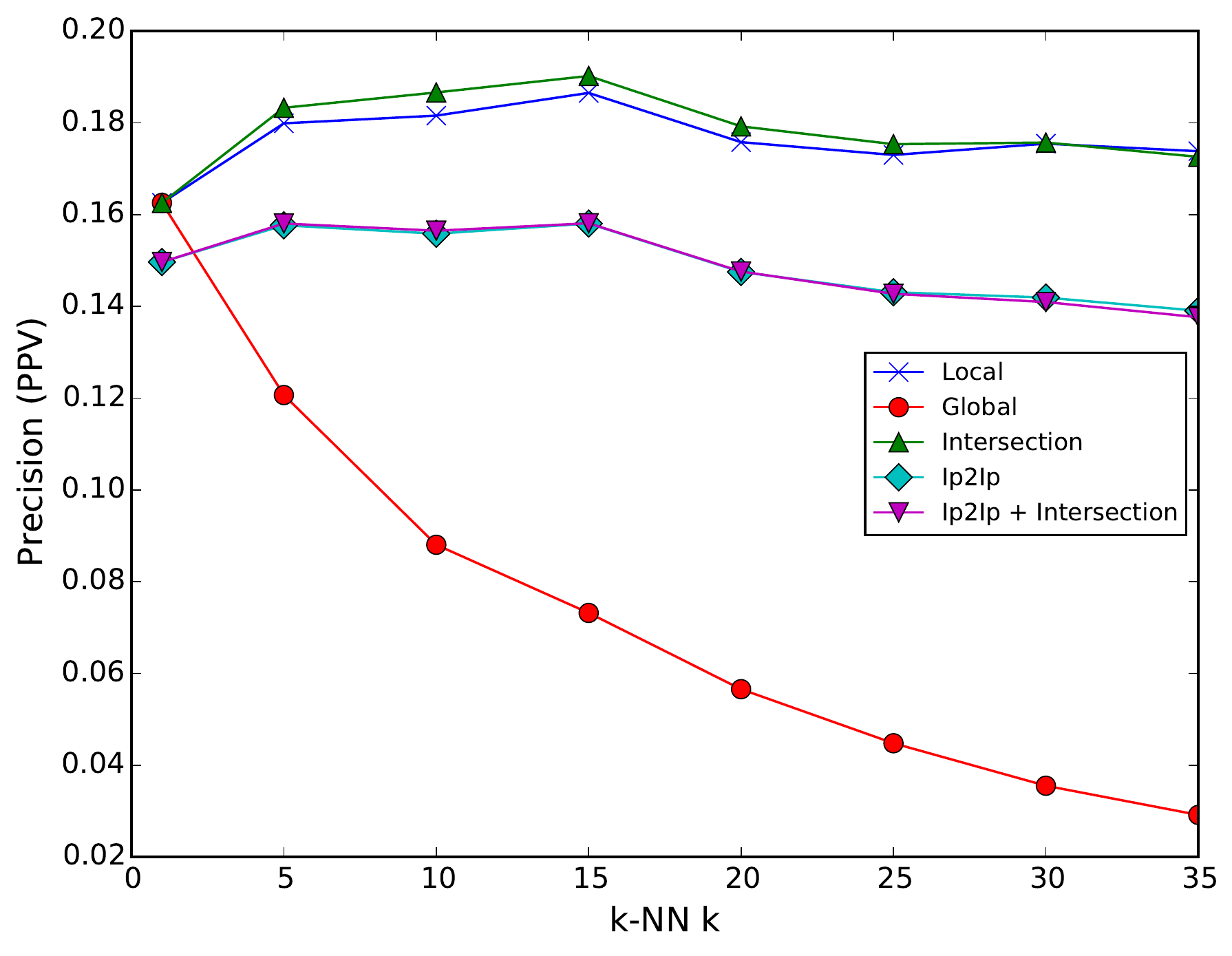}
\caption{\label{fig:knn-ppv}}
\end{subfigure}
\begin{subfigure}[t]{0.325\textwidth}
\centering
\includegraphics[width=1\textwidth,height=0.18\textheight]{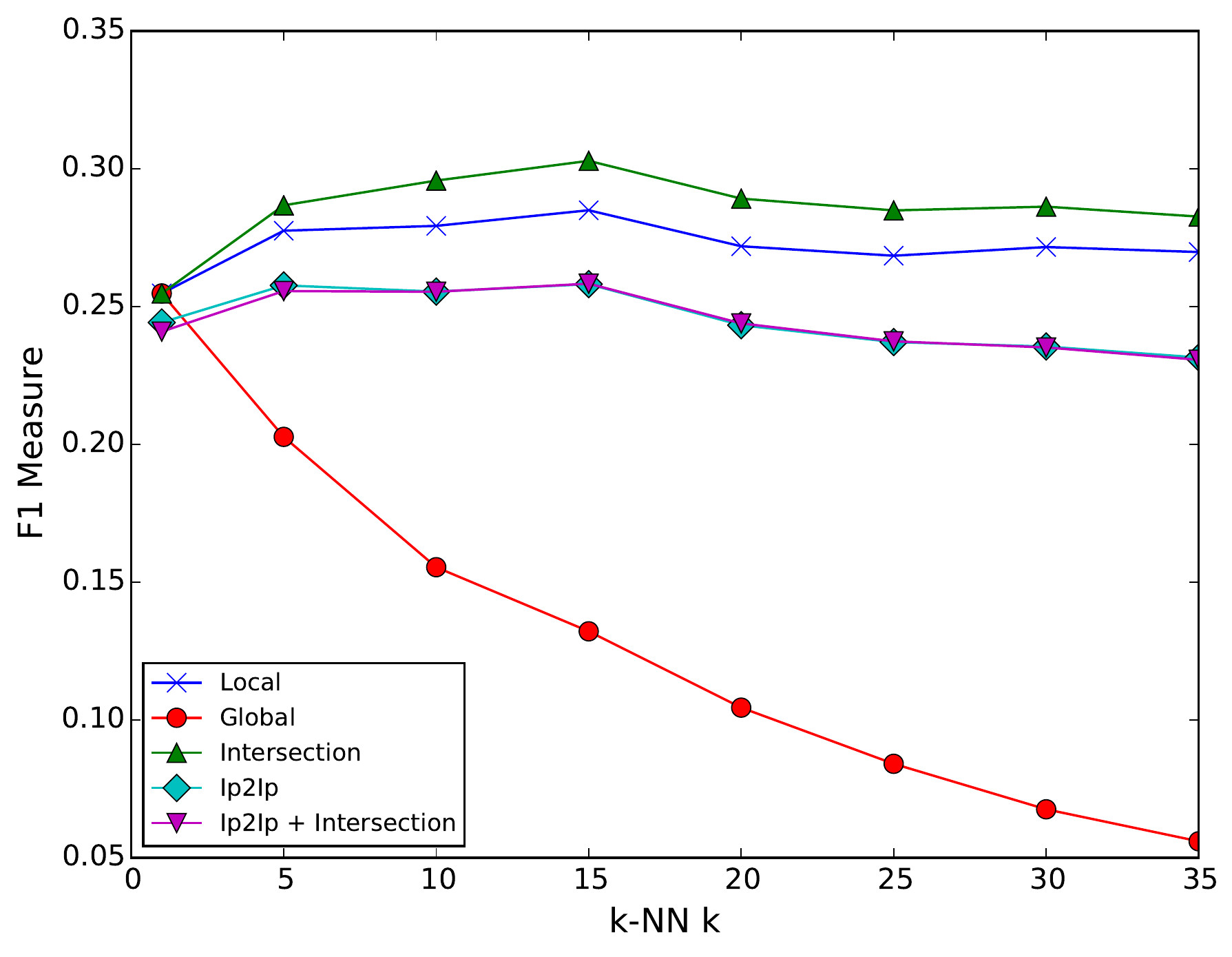}
\caption{\label{fig:knn-f1}}
\end{subfigure}
\vspace{-0.25cm}
\caption{k-NN: (a) TP improvement, (b) FP increase (y-axis in log scale), (c) FN increase, (d) TPR, (e) Precision, (f) F1 measure.} %
\vspace{-0.2cm}
\label{fig:knn}
\end{figure*}

\descr{Settings.} We once again use datasets and settings from Section~\ref{sec:preliminaries}.
Also, for the \iptip method, we only consider the top-1000 attackers (i.e., the top-1000 \textit{heavy hitters}) in each cluster, for each 5-day training-set window, rather than looking for correlations over all the /24 IP space. 
We fix the $k$ value for the k-NN based recommendation to 50, as it provides the best results in our experiments.

\descr{k-means.}
Next, we use k-means %
for clustering and %
decide to restrict to {\em stronger correlations}, 
by only taking into account organizations closer to the cluster's centroid, and excluding the rest of them as outliers. 
We set a distance threshold and \edit{choose the value that yields the best result, i.e., the cluster distance value below which 40\% of the organizations can be found}.
Fig.~\ref{fig:kmeans-tp-incr}--\ref{fig:kmeans-fn-incr} plot the average improvement in TP and increase in FP and FN. $TP_{impr}$ is almost constant with \iptip ($0.2$) independent of the cluster sizes, while with the other methods it decreases faster due to the distance thresholds, ranging from $1.1$ with global for $k=1$ to $0.1$ of intersection for $k=35$. %
\iptip shows steady $FN_{incr}$ values compared to other methods ($-0.2$, i.e., a $20\%$ decrease) which leads to a better performance in TPR, as shown in Fig.~\ref{fig:kmeans-tpr}, for $k \geq 10$ (up to $0.71$).
Furthermore, intersection yields the best performance in $FN_{incr}$ ($-0.52$), with $k=1$.
Fig.~\ref{fig:kmeans-f1} shows the best F1 measure ($0.30$) is reached with $k=5$, due to a peak both in PPV and TPR. \iptip performs slightly worse ($0.23$) than local ($0.28$) while poor F1 values for global, with $k=35$, ($0.18$) are due to its bad PPV ($0.10$) -- see Fig.~\ref{fig:kmeans-ppv}.

\descr{k-NN.} Recall that $k$ indicates the number of nearest neighbors that each entity considers as its most similar ones. Thus, organizations can end up in more than one neighborhood. %
Since the algorithm builds a neighborhood for each organization, 
not all clusters have the same \textit{strength}, so 
we only consider \textit{strong} clusters in terms of their members similarity and as done with k-means, after tuning the parameters, we set a distance threshold as the 40th percentile to leave possible outliers out of the clusters. 
From Fig.~\ref{fig:knn-tp-increase}, we observe that \iptiplus yields the second best performance in $TP_{impr}$ ($0.38$, with $k=35$), while global peaks at $0.60$. 
In terms of $FP_{incr}$, \iptip doubles it (for $k=35$), while intersection achieves the lowest value with $0.51$ (again, for $k=35$).
As with previous clustering algorithms, we notice that intersection yields the best decrease in FN, i.e., $-0.5$ with $k=35$. 
Intersection also achieves the highest TPR (up to $0.77$) with larger cluster sizes (i.e., for $k \geq 10)$, while its combination with the \iptip reduces it ($0.71$) -- see Fig.~\ref{fig:knn-tpr}. 
Fig.~\ref{fig:knn-ppv} shows that intersection has the best PPV ($0.19$ for $k=15$), similar to local ($0.18$), while \iptip performs worse ($0.16$) due to higher $FP_{incr}$ (almost doubling the FP for $k=35$).
Finally, from Fig.~\ref{fig:knn-f1}, note that intersection yields the highest F1 ($0.30$ for $k=15$).

\descr{Summary of results.} We summarize the best results for each clustering algorithm, in terms of best F1, recall, precision, and $TP_{impr}$ in Tables~\ref{table:f1-best}--\ref{table:tp-impr-best}. 
We note that intersection is that sharing mechanism
that maximizes all metrics, except for $TP_{impr}$, which is instead maximized with \iptiplus.
Both k-means and k-NN peak at $0.30$ in F1 including, respectively, $280$ and $240$ collaborators over all time windows. 
Agglomerative clustering involves all $700$ contributors and achieves $F1=0.27$. 
k-NN with $k=35$ yields the best results for TPR ($0.77$), while both k-NN with $k=35$ and k-means with $k=5$ achieve $0.19$ in PPV.
In terms of $TP_{impr}$, k-means reaches a maximum of $0.61$ with $k=1$ and clusters of size $28$ on average, selecting $270$ collaborators overall. Slightly lower improvements are achieved with other clustering algorithms, but with more collaborators benefiting from sharing, as well as fewer FP.

Data sharing always helps organizations forecast attacks, compared to performing predictions locally. Predicting based on all data from collaborators yields the highest improvement in $TP_{impr}$ -- especially for bigger clusters -- but with a dramatic increase in $FP_{incr}$. 
When organizations share correlated attacks (\iptip), we observe a steady $TP_{impr}$, while sharing common attacks (intersection) outperforms the former when bigger clusters are formed. However, intersection introduces lower $FP_{incr}$, ultimately leading to better precision and F1 measures.
\iptiplus always outperforms 
the two separate methods in terms of $TP_{impr}$, thus, it is the recommended strategy if one only wants to maximize the number of predicted attacks. %

\begin{table}[!t]
\centering
\begin{minipage}[t]{0.99\columnwidth}

\centering%
\resizebox{1.02\columnwidth}{!}{%
\hspace*{-0.2cm}
\begin{tabular}{| c | c | c | c | c | c | c | c | c | c |}
\hline
\multicolumn{2}{|c|}{{\bf Setting}} & \multicolumn{8}{c|}{\bf Max F1 [Sharing Intersection]}  \\
\hline
Clustering & $k$ & \scalebox{.9}[1.0]{Avg Size} & \#Coll. & TPR & PPV & $TP_{impr}$ & $FP_{incr}$ & $FN_{incr}$ & $~F1~$  \\
\hline
Agglom. & ~15~  & 4.6 & 700 & 0.72 & 0.16  & $0.38 \pm 3.51$ & $0.71 \pm 4.49$ & -0.42 & 0.27 \\ \hline
k-means & 5  & 5.8 & 280 & 0.73 & 0.19 & $0.44 \pm 4.21 $ & $0.31 \pm 1.47$ & -0.32 & 0.30 \\ \hline
k-NN & 15  & 6 & 240 & 0.74 & 0.19 & $0.27 \pm 0.20$ & $0.33 \pm 0.28$ &-0.37 & 0.30 \\ \hline
\end{tabular}}
\vspace{-0.3cm}
\caption{Best Cases of our Experiments for F1.}
\label{table:f1-best}
\vspace{0.1cm}

\centering%
\resizebox{1.02\columnwidth}{!}{%
\hspace*{-0.2cm}
\begin{tabular}{| c | c | c | c | c | c | c | c | c | c | }
\hline
\multicolumn{2}{|c|}{{\bf Setting}} & \multicolumn{8}{c|}{\bf Max TPR [Sharing Intersection]}  \\
\hline
Clustering & $k$ & \scalebox{.9}[1.0]{Avg Size} & \#Coll. & TPR & PPV & $TP_{impr}$ & $FP_{incr}$ & $FN_{incr}$ & $~F1$  \\
\hline
Agglom. & 1  & 70 & 700 & 0.76 & 0.15 & $0.50 \pm 3.95$ & $1.12 \pm 6.98$ & -0.53 & 0.25 \\ \hline
k-means & 5  & 5.8 & 280 & 0.73 & 0.19 & $0.44 \pm 4.21$ & $0.31 \pm 1.47$ & -0.32 & 0.30\\ \hline
k-NN & ~35~ & 14 & 320 & 0.77 & 0.17 & $0.32 \pm 0.21$ & $0.51 \pm 0.50$ & -0.49 & 0.28 \\ \hline
\end{tabular}}
\vspace{-0.3cm}
\caption{Best Cases of our Experiments for TPR.}
\label{table:tpr-best}
\vspace{-0.2cm}

\end{minipage}
\end{table}

\descr{Impact of cluster size.} With agglomerative clustering, each organization is assigned to exactly one cluster and thus participates in/benefits from collaboration. We observe higher TPR for bigger clusters and, generally, a stable improvement in TP is achieved on average. 
Similar results are obtained with k-means when all organizations are assigned to clusters. However, when we set a distance threshold, creating more consistent clusters, we observe fluctuations in TPR: as clusters get smaller much faster (in relation to $k$ value), \iptip starts outperforming intersection. This indicates that correlated attacks can improve knowledge of organizations and enhance their local predictions, especially in smaller clusters. %
With k-NN, a different behavior is observed: for smaller clusters, \iptip achieves higher TPR (up to $0.7$ for $k=5$) but, as clusters get bigger, intersection yields the best results (up to $0.77$ for $k=35$). 
Overall, collaborating in big clusters leads to high $TP_{impr}$ but at the same time it introduces significant $FP_{incr}$.

\descr{Increase/Improvement in TP/FP/FN.}
We observe that, for all clustering algorithms, maximizing $TP_{impr}$ always leads to higher $FP_{incr}$, from $1.51$ of k-NN up to $5.33$ of Agglomerative.
The settings that maximize the F1 measure, TPR, and PPV, (when sharing intersection) also minimize $FN_{incr}$, e.g. agglomerative with $k=1$ achieves $-0.53$ $FN_{incr}$. 
In general, we observe that (privacy-friendly) collaboration does yield a remarkable increase in TP but also in FP, which results in a limited improvement in F1 score compared to predicting using local logs only. 

\edit{Overall, our measurements allow us to quantify how different collaboration strategies affect prediction in terms of increasing true positives, false positive, and false negatives, and in general precision, recall, and F1. Ultimately, the main goal is to find settings that improve TP while keeping the increase in FP as low as possible. In this context, the best approach is sharing common and correlated attacks (\iptiplus) with k-NN (see Table~\ref{table:tp-impr-best}).}

\descr{Hybrid approach vs state of the art~\cite{freudiger2015controlled,soldo2010predictive}.}
When comparing the hybrid approach to Soldo et al.~\cite{soldo2010predictive}, %
we observe that~\cite{soldo2010predictive} achieves higher maximum
$TP_{impr}$ ($0.99$ vs $0.61$ with k-means, $k=1$). However, our privacy-preserving techniques outperform~\cite{soldo2010predictive} in terms of recall (TPR) (e.g., with k-NN we reach $0.77$ %
compared to their $0.66$, i.e. up to 18\% increase) as well as precision ($0.19$ with k-means, $k=5$ vs $0.08$, i.e. up to 15\% increase) and F1 measure ($0.30$ with k-NN, $k=15$ vs $0.14$). 
Finally, comparing the hybrid approach to~\cite{freudiger2015controlled}%
the former yields better results in terms of $TP_{impr}$ (0.61 for k-means, $k=1$ vs 0.13 for top 3\% of global pairs) and TPR (0.77 for k-NN, $k=35$ vs 0.66 for top 1\% of global pairs),
but similar F1 score (0.30 vs 0.28), due to the latter's smaller increase in FP. 

\edit{Overall, we conclude that a controlled data sharing approach, compared to a centralized one, helps organizations find a better trade-off between prediction improvement and increase in false positives, while minimizing exposure of possibly confidential data.}

\begin{table}[!t]
\centering
\begin{minipage}[t]{0.99\columnwidth}

\centering%
\resizebox{1.02\columnwidth}{!}{%
\hspace*{-0.2cm}
\begin{tabular}{| c | c | c | c | c | c | c | c | c | c | }
\hline
\multicolumn{2}{|c|}{{\bf Setting}} & \multicolumn{8}{c|}{\bf Max PPV [Sharing Intersection]}  \\
\hline
Clustering & $k$ & \scalebox{.9}[1.0]{Avg Size} & \#Coll. & TPR & PPV & $TP_{impr}$ & $FP_{incr}$ & $FN_{incr}$ & $~F1$  \\
\hline
Agglom. & 25 & 2.8 & 700 & 0.69 & 0.16 & $0.33\pm 3.29$ & $0.47 \pm 2.23$ & -0.35 & 0.26 \\ \hline
k-means & 5  & 5.8 & 280 & 0.73 & 0.19 & $0.44 \pm 4.21$ & $0.31 \pm 1.47$ & -0.32 & 0.30\\ \hline
k-NN & 15  & 6 & 240 & 0.74 & 0.19 & $0.27 \pm 0.20$ & $0.33 \pm 0.28$ & -0.37 & 0.30 \\ \hline
\end{tabular}}
\vspace{-0.3cm}
\caption{Best Cases of our Experiments for PPV.}
\label{table:ppv-best}
\vspace{0.1cm}

\centering%
\resizebox{1.02\columnwidth}{!}{%
\hspace*{-0.2cm}
\begin{tabular}{| c | c | c | c | c | c | c | c | c | c |}
\hline
\multicolumn{2}{|c|}{{\bf Setting}} & \multicolumn{8}{c|}{\bf Max TP Improvement [Sharing \iptiplus]}  \\
\hline
Clustering & $k$ & \scalebox{.9}[1.0]{Avg Size} & \#Coll. & TPR & PPV & $TP_{impr}$ & $FP_{incr}$ & $FN_{incr} $ & $~F1~$  \\
\hline
Agglom. & 1  & 70 & 700 & 0.67 & 0.11 & $0.52 \pm 3.95$ & $5.33 \pm 16.9$ & -0.08 & 0.19 \\ \hline
k-means & 1 & 28 & 270 & 0.64 & 0.11 & $0.61 \pm 5.36$ & $3.55 \pm 7.17$ & -0.17 & 0.18 \\ \hline
k-NN & ~35~ & 14 & 320 & 0.71 & 0.14 & $0.38 \pm 0.25$ & $1.51 \pm 1.02$ & -0.19 & 0.23 \\ \hline
\end{tabular}}
\vspace{-0.3cm}
\caption{Best Cases of our Experiments for $TP_{impr}$.}
\label{table:tp-impr-best}
\vspace*{-0.3cm}
\end{minipage}
\end{table}

\section{Implementing At Scale}
\label{section:extension}
As discussed above, our hybrid system involves four steps: (1) secure computation of pairwise similarity, (2) clustering, (3) secure data sharing within the clusters, and (4) time-series prediction. 
To assess its scalability, we need to evaluate computation and communication complexities incurred by each step. Naturally, steps (1) and (3) dominate complexities as they require running a number of cryptographic operations (involving public-key crypto) that depends on the number of organizations involved.
In fact, clustering incurs a negligible overhead: on commodity hardware, to perform clustering with 1,000 organizations, it takes $6.1ms$ for k-means, $81ms$ for agglomerative and $5.2ms$ for k-NN ($k=2$).%
Also, time-series EWMA prediction requires $4.6\mu s$ per IP, 
so it takes $4.6ms$ for 1,000 IPs.
As we compute pairwise similarity based on the amount of common attacks between two organizations, and support its secure computation via PSI-CA~\cite{de2012fast}, step (1) requires a number of protocol runs {\em quadratic} in the number of organizations. In our experiments, it takes $1.98s$ and $2.12$MB bandwidth for one protocol execution, using 2048-bit moduli, with sets of size 4,000 (the average number of attacks observed by each organization).
As for (3), i.e., secure within-cluster sharing of events related to common attacks (intersection), we rely on PSI-DT~\cite{de2010practical}, and it takes $1.24s$ and $2.18$MB for a single execution with the same settings. 
Therefore, complexities may become prohibitive when more organizations are involved or more alerts are used.

\begin{figure}[t]
\centering
\begin{minipage}[t]{1\columnwidth}
\begin{algorithm}[H]
\caption{\small{\sc Encryption} [All Organizations]}
\label{alg:encryption}
\begin{algorithmic}\small
\FOR{\textbf{each} $O_i \in \mathcal{O}$ }
	\STATE $S_{i} \gets \emptyset$, $E_{i} \gets \emptyset$, $K_{i} \gets \emptyset $ 
	\FOR{\textbf{each} $(d_j,time_j) \in D_i$}
		\FOR{$cnt := 1$ \textbf{to} $COUNT(d_j) $}  
			\STATE $S_{i} \gets S_{i} \cup \mbox{PRP}_{k}(d_j || cnt)$ 
			\STATE $k_{j} \gets {H(d_j || cnt)}$
			\STATE $E_{i} \gets E_{i} \cup \mbox{Enc}_{k_j}(d_j, time_{j})$ 
			\STATE $K_{i} \gets K_{i} \cup k_{j}$
		\ENDFOR
	\ENDFOR	
	\STATE Send $S_{i}, E_{i}$ to $\STA$ and store $K_{i}$
\ENDFOR
\end{algorithmic}
\end{algorithm}
\end{minipage}
\\
\begin{minipage}[t]{1\columnwidth}
\begin{algorithm}[H]
\caption{\small{\sc \oto Computation} [STA]}
\label{alg:o2o}
\begin{algorithmic}\small
\FOR{\textbf{each} $O_{i} \in \mathcal{O}$ }
\FOR{\textbf{each} $O_{j} \neq O_{i}$ }
\STATE $\mbox{\oto}[i,j] \gets | S_{i} \cap S_{j} | $
\STATE $\mbox{Buff}[i,j] \gets \{(\ell, E_{j_\ell} ),  \forall \ell \in \mbox{INDEX}(S_i \cap S_j)\}$
\ENDFOR
\ENDFOR
\STATE Perform Clustering on O2O$[\cdot,\cdot]$
\STATE Send relevant Buff$[\cdot,\cdot]$ entries to organizations in the same cluster 
\end{algorithmic}
\end{algorithm}
\end{minipage}
\\
\begin{minipage}[t]{1\columnwidth}
\begin{algorithm}[H]
\caption{\small{\sc Log Sharing} [Organizations in  $C^*$]}
\label{alg:shar}
\begin{algorithmic}\small
\FOR{\textbf{each} $O_{i} \in C^*$ }
\STATE $S'_{i} \gets \emptyset$
\FOR{\textbf{each} $O_{j} \neq O_{i} \in C^*$ }
\FOR{\textbf{each} $(\ell, E_{j_\ell}) \in \mbox{Buff}[i, j] $ }
\STATE $ S'_{i} = S'_{i} \cup \mbox{Dec}_{k_{\ell}}(E_{\ell}) $
\ENDFOR
\ENDFOR
\ENDFOR
\end{algorithmic}
\end{algorithm}
\end{minipage}
\vspace{-0.35cm}
\end{figure}

\begin{figure*}[!t]
\centering
\begin{subfigure}[t]{0.33\textwidth}
\centering
\includegraphics[width=1\textwidth,height=0.18\textheight]{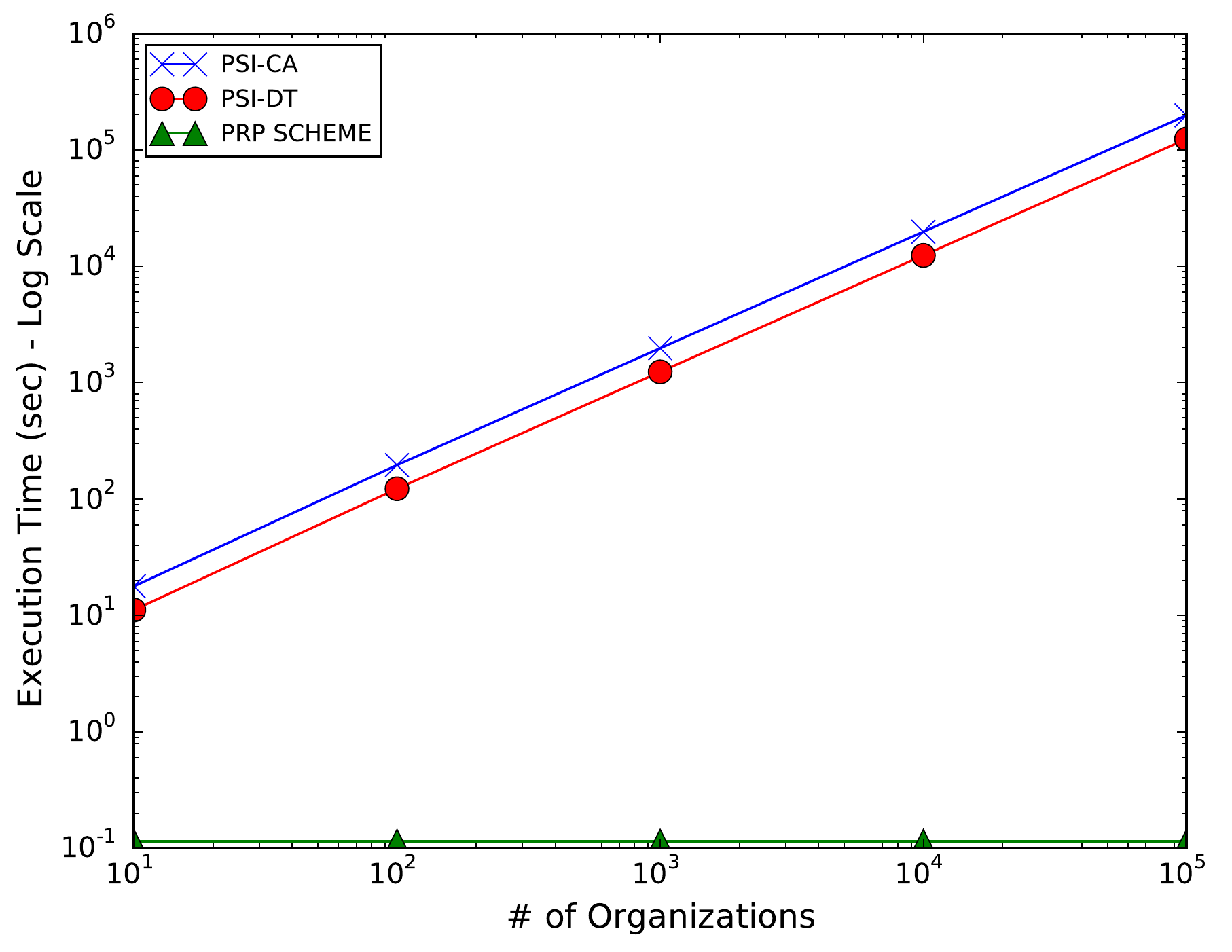}
\vspace{-0.5cm}
\caption{\label{fig:computation}}
\end{subfigure}
\begin{subfigure}[t]{0.33\textwidth}
\centering
\includegraphics[width=1\textwidth,height=0.18\textheight]{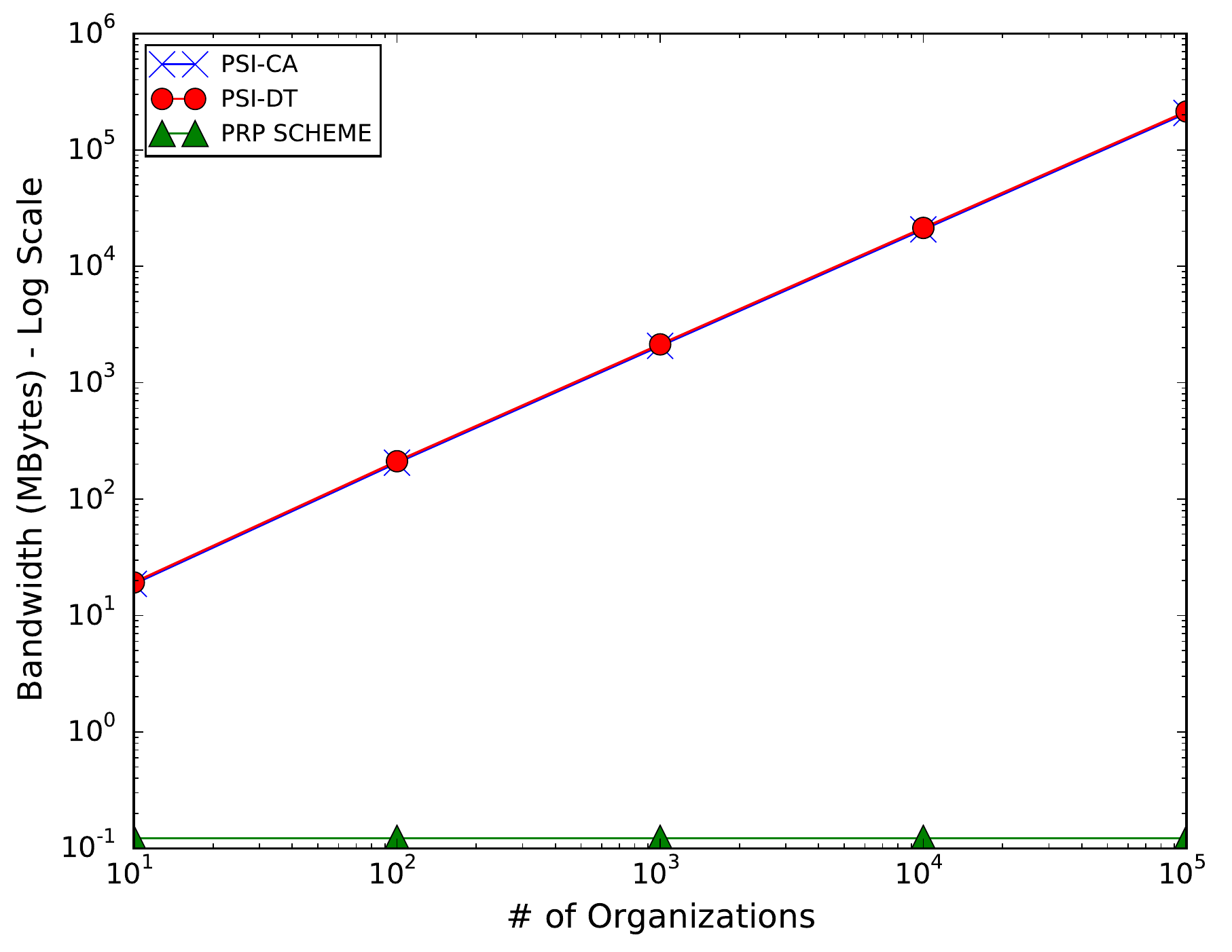}
\vspace{-0.5cm}
\caption{\label{fig:communication}}
\end{subfigure}
\begin{subfigure}[t]{0.33\textwidth}
\centering
\includegraphics[width=1\textwidth,height=0.18\textheight]{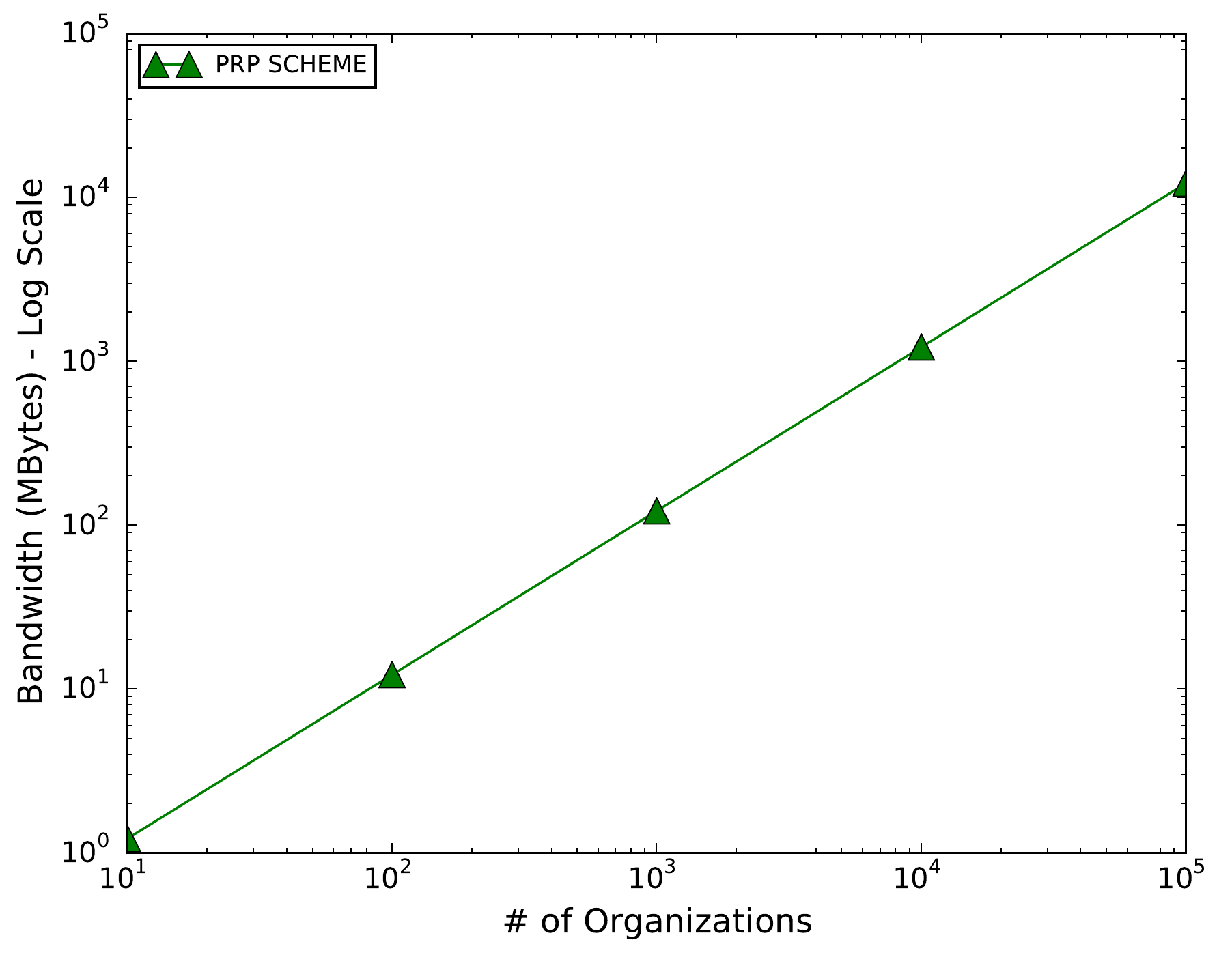}
\vspace{-0.5cm}
\caption{\label{fig:sta_communication}}
\end{subfigure}
\vspace{-0.3cm}
\caption{Computation (a) and communication (b) overhead at each organization for PSI-CA, PSI-DT, and PRP-based scheme, and  communication overhead at the \STA in PRP scheme (c).}
\vspace{-0.15cm}
\end{figure*}

Aiming to improve scalability, we also implement a variant supporting secure computation of pairwise similarity as well as secure log sharing {\em without} a quadratic number of public-key operations/quadratic communication overhead.
Recall that we rely on a semi-trusted authority, \STA, for clustering and coordination, which is assumed to follow protocol specifications and not to collude with other organizations, thus, we can actually use it to also help with secure computations. 
Inspired by Kamara et al.'s server-aided PSI~\cite{kamara2014scaling}, we extend our framework by replacing public-key cryptography operations with pseudo-random permutations (PRP), which we instantiate using AES.
Specifically, we minimize interactions among pairs of organizations so that the complexity incurred by each of them is constant, while only imposing a minimal, linear communication overhead on \STA.

Our extension involves four phases: (i) {\em setup} %
where, as in~\cite{kamara2014scaling}, one organization generates a random key $\mathcal{\kappa}$ and sends it to the other organizations, (ii) {\em encryption}  (see Algorithm~\ref{alg:encryption}),
where each organization $O_{i}$ evaluates the PRP on each entry $d_{j}$ in their sets and encrypts the associated timestamp $time_{j}$, (iii) {\em \oto computation}  (see Algorithm~\ref{alg:o2o}), 
where \STA computes the magnitude of common attacks between each pair of organizations in order to perform clustering, and (iv) {\em log sharing} (see Algorithm~\ref{alg:shar}), 
where organizations in the same cluster $C^{*}$ receive information about common attacks ($S'_{i}$-s).
Note that building the \oto matrix is actually optimized using hash tables (i.e., \textit{dense\_hash\_set} and \textit{dense\_hash\_map} from Sparehash. 
Also, since sets in our system are multi-sets, we concatenate counters to the IP address, so that the \STA cannot tell which and how many IPs appear more than once.

\descr{Experimental Evaluation.}
We benchmark the performance of PSI-CA~\cite{de2010practical} and PSI-DT~\cite{de2010practical} using 2048-bit moduli,
modifying the OpenSSL/GMP-based C implementation in~\cite{de2012experimenting},
as well as the PRP-based scheme presented above and inspired by 
Kamara et al.'s work~\cite{kamara2014scaling}. Experiments are run 
using two 2.3GHz Intel Core i5 CPUs with 8GB of RAM connected via a $100$Mbps Ethernet link.
Figures~\ref{fig:computation} and~\ref{fig:communication} plot computation and communication complexities incurred by an individual organization vis-\`a-vis the total number of organizations involved in the system, while Fig.~\ref{fig:sta_communication} reports the communication overhead introduced on the \STA-side for the PRP scheme.
As expected, complexities for PSI-CA/PSI-DT protocols on each organization %
grow linearly in the number of organizations (hence, quadratic overall). For instance, if 1,000 organizations are involved, it would take about 16 minutes per organization, each transmitting 1GB.
Whereas, the PRP-based scheme incurs constant complexities on each organization ($57.6ms$ and $120$KB) and a low
communication overhead on the \STA (about 100MB) for 1,000 organizations (Fig.~\ref{fig:sta_communication}).

We also evaluate the \iptip method whereby organizations interact with \STA in order to discover cluster-wide correlated attacks. %
Assuming clusters of $100$ organizations and an \iptip matrix of $(2^{24}\cdot 2^{24})/2$ (recall  we consider the whole /24 IP space), we measure a $2.7s$ running time per organization with $41$KB of bandwidth as well as a $0.07s$ overhead on the \STA with $4.1$MB bandwidth. 
Using the private Count-Min sketch based implementation by
Melis et al.~\cite{melis}, we can compress to a logarithmic factor with a small, bounded loss, and 
the private aggregation is done over 10,336 elements.
Even if clusters are bigger than 100, as detailed in~\cite{melis}, one can still perform private aggregation on multiple subgroups (e.g., of size 100) without endangering organizations' privacy.

\descr{Security.} Protocols do not leak {\em any} information about the logs of each organization to the \STA, with or without using the server-aided variant. Clustering is performed over similarity measures computed obliviously to \STA, and so does within-cluster data sharing. Privacy-preserving computation occurs by using existing secure protocols such as PSI-CA/PSI-DT by De Cristofaro et al.~\cite{de2010practical,de2012fast}), server-aided PSI by Kamara et al.~\cite{kamara2014scaling}, as well as private recommendation via succinct sketches by Melis et al.~\cite{melis}. Therefore, we do not provide any additional proofs in the paper as the security of our techniques straightforwardly relies on that of these protocols.

\section{Conclusion}\label{sec:discussion}

This paper presented the result of a measurement study of collaborative predictive blacklisting (CPB). 
We evaluated a number of metrics on a real-world dataset obtained from DShield.org, 
aiming to shed light on the effects of collaboration when considering
two state-of-the-art approaches, one, non privacy-preserving, relying on trusted central party~\cite{soldo2010predictive} and another peer-to-peer using privacy-preserving data sharing~\cite{freudiger2015controlled}. We also introduced a third, hybrid approach that aims to combine the best of the two worlds. 

\edit{Naturally, having access to more logs does not necessarily result in better predictions.} In fact, our experiments showed that the techniques proposed by Soldo et al.~\cite{soldo2010predictive} achieve impressive hit counts (almost doubling the number of correct predictions compared to local predictions) but suffer from poor precision due to high FP. On the other hand, the privacy-friendly decentralized system proposed by Freudiger et al.~\cite{freudiger2015controlled} achieves better F1 scores overall, although with a decreased improvement in TP. Finally, our analysis shows that our hybrid approach outperforms both approaches, balancing out true and false positives, while maintaining privacy protection. 

As part of future work, we plan to conduct a longitudinal measurement to fully grasp the effectiveness of privacy-enhanced CPB in the wild, 
\edit{apply our methods to other datasets, and experiment with more advanced machine learning techniques to improve overall performances.}

\bibliographystyle{abbrv}
%\bibliography{bibfile}

\appendix

\section{Reproducing Our Experiments}

We provide detailed information for researchers wishing to reproduce the experimental results presented in this paper. %
Please install Python 2.7 as well as the following Python packages: numpy 1.14.0, scipy 1.0.0, scikit-learn 0.19.1, pandas 0.22.0 and matplotlib 2.0.0. All of the above packages can be installed via ``pip''. 

\descr{Code.} Source code is available at the following git repository:\\
\hspace*{0.5cm} \url{https://github.com/mex2meou/collsec.git}

\descr{Dataset.} To obtain the DShield dataset that was used in our experiments, use the following download link and extract its contents (i.e., the .pkl files) in the `data' folder of the cloned repository.

\begin{lstlisting}
wget https://www.dropbox.com/s/kmiejttl4ceufpp/data.zip
cp data.zip collsec/data
cd collsec/data
unzip data.zip
\end{lstlisting}

\descr{Soldo et al~\cite{soldo2010predictive}.} To replicate the experiments for Soldo et al~\cite{soldo2010predictive}'s implicit recommendation system, we also need the MATLAB implementation of Chakrabarti et al.~\cite{chakrabarti2004fully} for the Cross Associations (CA) co-clustering algorithm. To this end, one should install Octave 4.0.0 as well as the Python package oct2py 3.5.0 (which can also be installed via ``pip''). First, to compile the Cross Associations algorithm follow the steps:

\begin{lstlisting}
cd collsec/soldo/CA_python
octave
mex cc_col_nz.c
quit
\end{lstlisting}

Then, to link our Python implementation with the Octave workspace of CA configure accordingly the path in the file `collsec/soldo/CA\_python/ca\_utils.py' (line 6). Finally, to run the experiments of Section~\ref{sec:soldo_approach}:

\begin{lstlisting}
cd collsec/soldo
python soldo.py
\end{lstlisting}

\descrit{Note.} To configure the parameter $k$ of the k-NN algorithm included in the ensemble method of~\cite{soldo2010predictive} modify the file `collsec/soldo/top\_neighbors.py' (line 4). Moreover, if experiments for various values of $k$ are executed, modify the file `collsec/soldo/soldo.py' (line 41) to prevent the CA algorithm from running again.

\descr{Controlled Data Sharing.} To repeat the experiments for the Controlled Data Sharing system by Freudiger et al.~\cite{freudiger2015controlled}, i.e., Section~\ref{sec:freudiger}:

\begin{lstlisting}
%
cd collsec/dimva-global
python dimva-global.py

%
cd collsec/dimva-local
python dimva-local.py
\end{lstlisting}

\descrit{Note.} To configure the length of the training and testing windows for the system of Freudiger et al.~\cite{freudiger2015controlled}, modify the file `collsec/utils/dimva\_util.py'.

\descr{Hybrid Approach.} To launch the experiments for our proposed hybrid scheme (see Section~\ref{section:methodology}) please execute the following steps:

\begin{lstlisting}
%
%
cd collsec/agglomerative
python agglomerative.py

%
%
cd collsec/kmeans
python kmeans.py

%
%
cd collsec/knn
python knn.py
\end{lstlisting}
\descrit{Note.} Our implementation by default is configured to utilize a 5-day training window and a 1-day testing one as done in previous work~\cite{freudiger2015controlled,soldo2010predictive}. If one wants to change this setting, please adjust the parameters indicated in the files `collsec/utils/util.py' and `collsec/utils/time\_series.py'.
\vfill
\pagebreak

\descr{Results.} The results of all the above scripts are stored in the folder titled `collsec/results'. To visualize the results and obtain the figures presented in the paper, type the following commands:

\begin{lstlisting}
cd collsec/results

%
python soldo_plots.py

%
python dimva_global_plots.py

%
python dimva_local_plots.py

%
python hybrid_plots.py
\end{lstlisting}

\end{document}